\documentstyle[aps,prl,epsf,floats,multicol,amssymb,tighten]{revtex}

\begin{document}

\def\oppropto{\mathop{\propto}} 
\def\opsimeq{\mathop{\simeq}}
\def\opoverderline{\mathop{\overline}}
\def\operarrow{\mathop{\longrightarrow}}
\def\opsim{\mathop{\sim}} 

\newcommand{\fig}[2]{\epsfxsize=#1\epsfbox{#2}} \reversemarginpar 
\newcommand{\mnote}[1]{$^*$\marginpar{$^*$ {\footnotesize #1}}}

\bibliographystyle{prsty}

\title{Energy dynamics in the Sinai model} 

\author{C\'ecile Monthus}
\address{Service de Physique Th\'eorique, CEA/DSM Saclay,
Unit\'e de recherche associ\'ee au CNRS, \\
91191 Gif-sur-Yvette, France}

\author{Pierre Le Doussal}
\address{CNRS-Laboratoire de Physique Th\'eorique de l'Ecole 
Normale Sup\'erieure, 24 rue Lhomond, F-75231
Paris}

\maketitle
\begin{abstract}
We study the time dependent potential energy 
$W(t)=U(x(0)) - U(x(t))$ of a particle diffusing 
in a one dimensional random force field (the Sinai model).
Using the real space renormalization group method (RSRG),
we obtain the exact large time limit of the probability distribution 
of the scaling variable $w=W(t)/(T \ln t)$. This distribution 
exhibits a {\it nonanalytic} behaviour at $w=1$. These results
are extended to a small non-zero applied field. Using the constrained
path integral method, we moreover compute the joint distribution of energy 
$W(t)$ and position $x(t)$ at time $t$. In presence of a reflecting boundary
at the starting point, with possibly some drift in the + direction,
the RSRG very simply yields the one time and aging two-time behavior 
of this joint probability. It exhibits differences in behaviour
compared to the unbounded motion, such as analyticity.
Relations with some magnetization 
distributions in the 1D spin glass are discussed. 
\end{abstract}


\widetext

\newpage



\section{Introduction}

The description of glassy activated dynamics is one of the challenges of
the present theory of glasses and disordered systems \cite{review,frg}. In fully
connected, mean field approximations of these systems, 
the energy initially decreases and saturates at a threshold:
the asymptotic large time dynamics then remains entirely dominated by high saddles
in energy landscape and not by barriers. A recently studied
model \cite{dilute}, still mean field in character but 
which incorporates some effects of finite connectivity
exhibits an additional regime of late times with subthreshold
dynamics. It mimics an activated dynamics between lower and
lower energy states, expected in any realistic physical system,
leading to an interesting, slow but unbounded, decrease of the energy.

At the other end of the spectrum, the Sinai model \cite{sinai} 
of a particle diffusing in a one dimensional quenched random force
field is dominated by activated dynamics and, despite its simplified
character, is a non trivial example of a glass. It has the advantage of
being analytically tractable, most notably through the real space
renormalization group method (RSRG) \cite{us_prl,us_sinai}
and by various other methods
\cite{sinai,kesten,derrida_pomeau,ledou_1d,monthus_diffusion,laloux_pld_sinai}.
Its extensions
to many particles describe a variety of coarsening, domain wall growth 
and reaction diffusion models with disorder in one dimension, also
analytically tractable via the RSRG \cite{us_rfim,us_rd,us_golosov,us_toy}.
It is thus interesting to study, within the Sinai model, 
how the energy decreases as a function of time, as the particle jumps from
deeper to deeper wells.

The aim of this paper is to obtain exact results on the large time
behaviour of the  energy $E(t)$ as a function of time
of a particle in Sinai model. The energy lansdcape, denoted
$U(x)$, is itself a random walk with 
$\overline{ (U(x) - U(x'))^2 } \sim |x-x'|$ at large scales. 
Fixing the energy to be zero at initial time, and denoting $x(t)$ 
the trajectory of a single particle, it is defined as:
\begin{eqnarray}
E(t) = U(x(t)) - U(x(0)) = - W(t)
\label{defenergy}
\end{eqnarray}
i.e it is minus the loss of potential energy $W(t)$, the quantity 
on which we focus from now on. In the Sinai model this
quantity exhibits non trivial sample to sample fluctuations, even in 
the large time limit. In this paper, we will thus be interested in the
distribution $D_t(W)$ of $W$ over both thermal histories
and environments (assuming a uniform probability to start at any site).
To our knowledge, this probability distribution of the energy 
has not been computed previously.

We will first consider the unbiased diffusion and use the RSRG
approach (Section \ref{energysym}). This method 
consists in coarse-graining the energy landscape 
in a way that exactly preserves the long time dynamics \cite{us_prl,daniel}. 
One decimates iteratively the smallest
energy barrier in the system stopping when the Arrhenius time to surmount 
the smallest remaining barrier $\Gamma$ is of order the time scale of interest $t$,
with $\Gamma = T \ln t$. The large time dynamics becomes asymptotically
identical to the so called ''effective dynamics'' where the particle is
with probability one at the bottom of the valley in the renormalized landscape
containing the starting point. As we find the energy at this point is such
that $W(t) \sim T \ln t$, i e the decrease is only logarithmic in time (much slower than the decay
of the energy in the model studied in Ref \cite{dilute}).
The correct scaling variable, which exhibits a universal probability distribution
at large time is then:
\begin{eqnarray} 
w = \frac{W(t)}{T \ln t}
\end{eqnarray}
and below, we compute this universal large time distribution $D(w)$. The most notable
result, besides the simple, exponential form of this distribution, is that this
function is found to be {\it non analytic} at $w=1$. For any finite time it is of course
analytic, but in the large time limit the distribution of the 
{\it rescaled} quantity exhibits a jump in its second derivative. As discussed
below this can be traced to the decimation process in the RSRG, but since this
is a property which can be measured with no a priori knowledge of the physics of the system,
e.g. via a numerical simulation,
it provides a rather remarkable ''experimental'' signature that the glass fixed point describing 
this system is of the type exactly solvable by RSRG, i.e. characterized
by an infinitely broad distribution of barriers. 
A interesting question 
is whether this is a more general property: this
suggests to search for observables exhibiting similar
signatures in other
systems with ''infinite disorder'' glass fixed point \cite{daniel}. 

The decrease with time of the energy can be related to the
$H$-theorem for this system. One recalls that the generalized free
energy:
\begin{eqnarray}
G(t) = \int dx [ T P(x,t) \ln P(x,t) + U(x) P(x,t) ]
\label{g}
\end{eqnarray}
is always decreasing in each sample :
\begin{eqnarray}
\partial_t G(t) = - \int dx \frac{J(x,t)^2}{P(x,t)} 
\label{dg}
\end{eqnarray}
a simple consequence of the Fokker Planck equation for the
probability distribution $P(x,t)$ of the position of the particle
$\partial_t P= - \partial_x J$, with $J=-T \partial_x P - P \partial_x U$.
Here one can easily see, since the effective dynamics becomes
asymptotically exact, that:
\begin{eqnarray}
\frac{G(t) }{\ln t } \opsimeq_{t \to \infty}   - \frac{ W(t)} { \ln t } 
\end{eqnarray}
confirming the considerations given in \cite{bounds}. 

Next, we will study the case of a small additional applied external 
force, for which the RSRG is still exact, and the result for
$D(W)$ will be generalized to that case in Section \ref{energybias}. In the following Section
\ref{pathjoint} we use a different and complementary method, the constrained
path integrals, to obtain the joint distribution of the position $x(t)$ and
energy $W(t)$. This yields information on
the distribution of positions corresponding to a fixed 
potential energy loss, and we show, in particular that this
distribution becomes deterministic at large $W$. 

Another situation studied here in Section \ref{reflecting} 
corresponds to the geometry of the half-line,
with a reflecting wall at the starting point. Then the RSRG allows to
solve not only the joint distribution of position and energy at time $t$ but
also the full aging two-time behavior of this joint probability.
The resulting physics is slightly different, however, as the
distribution of energy does not exhibit the nonanalytic behaviour.

Finally, the quantities studied here also have an interpretation in
terms of a spin model, the 1D spin glass. We show in Section \ref{sg} that the
distributions of magnetizations of that model on some intervals
is related to the above results.

\section{Distribution of energy }

\label{energyrsrg}

In this section, we compute the distribution $D_t(W)$ 
of the energy $W(t)$ (\ref{defenergy})
by solving the appropriate RSRG equation.

\subsection{Review of the main results from the real space renormalization method}

We now briefly recall the RSRG method for Sinai landscape,
and we refer the reader to \cite{us_sinai} for details.
In the RSRG, the original energy landscape is replaced by a renormalized
landscape at scale $\Gamma$, via an iterative decimation procedure where 
barriers smaller than $\Gamma$ are eliminated. This generates
a joint distribution $P_\Gamma(F,l)$ of the barrier $F$ and
length $l$ for the renormalized bonds in the
landscape at scale $\Gamma$.

In the symmetric case, the probability distribution of the
 rescaled variables $\eta= \frac{F-\Gamma}{\Gamma}$
and $\lambda=\frac{l}{\Gamma^2}$ flows towards the fixed-point \cite{us_sinai}
\begin{eqnarray}
P^*(\eta,\lambda)= LT^{-1}_{s \to \lambda}  \left( \frac{\sqrt s}{ \sinh \sqrt s } e^{- \eta \sqrt s \coth \sqrt s }\right)
\label{solu-sym}
\end{eqnarray}
In particular, the distribution of the barrier alone is a simple exponential
\begin{eqnarray}
P^*(\eta)= \int_0^{+\infty} d\lambda P^*(\eta,\lambda)= e^{-\eta} 
\label{etasym}
\end{eqnarray}

In the effective dynamics \cite{us_sinai}, 
the particle starting at $x=0$ at $t=0$
is at time $t$ in a finite neighborhood \cite{us_golosov} 
around the bottom of the  
bond of the renormalized lanscape at scale 
\begin{eqnarray}
\Gamma = T \ln t
\end{eqnarray}
 which contains the starting point $x=0$. 
After rescaling $X=\frac{x}{\Gamma^2}$, the effective dynamics
gives the asymptotic exact result of the Kesten distribution \cite{us_sinai}
for the rescaled position (diffusion front). Similarly here, the effective dynamics will give
the asymptotic exact result in the limit $t \to \infty$
for the rescaled variable of energy
\begin{eqnarray}
w = \frac{W}{ \Gamma} = \frac{W}{ T \ln t}
\label{defw}
\end{eqnarray}

\subsection{Distribution of energy for the symmetric Sinai diffusion}

\label{energysym}

We introduce the probability $D_{\Gamma}(F,W)$
that the point $x=0$ belongs at scale $\Gamma$
to a bond of barrier $F$ and is at potential $W$ above the lower potential 
of the bond, i.e. $U(x=0)-U_{min}=W$. It is normalized as:
\begin{eqnarray} 
\int_0^{F} dW D_{\Gamma}(F,W)=
\frac{1}{\overline{l_{\Gamma}}} 
\int_0^{\infty} dl l P_{\Gamma} (F,l)
\end{eqnarray}
since the probability that the origin $x=0$ belongs to
a renormalized bond of barrier $F$ and length $l$ is
$l P_\Gamma(F,l)/{\overline{l_{\Gamma}}}$. Within the 
effective dynamics, we may compute the probability distribution $D_t(W)$ 
of the energy $W$ at time $t$ as
\begin{eqnarray} 
D_t(W) = 
\int_{\max(\Gamma, W)}^{+\infty} dF D_{\Gamma}(F,W)
\end{eqnarray}

\begin{figure}[thb]

\centerline{\fig{14cm}{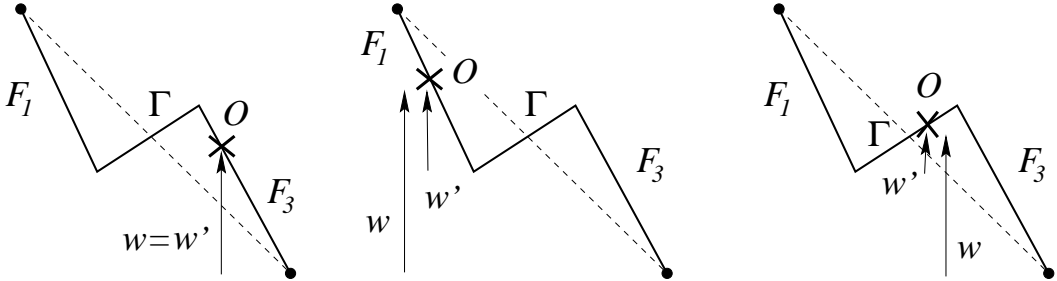}} 

\vskip 1cm

\caption{  Decimation of a bond of barrier $\Gamma$ that gives a renormalized bond of barrier $F=F_1+F_3-\Gamma$ : the point $O$ can be either on the bond $F_3$ that doesn't change the energy $W=W'$,
 or on the bonds $F_1$ or $\Gamma$ that both lead
to the incrementation $W=W'+F_3-\Gamma$ of the energy.\label{fig3} } 

\end{figure}

The consideration
of the various effects of the decimations processes (see figure \ref{fig3})
leads to the following RG equation for $D_{\Gamma}(F,W)$
\begin{eqnarray} 
\label{rgdg}
&&  \partial_{\Gamma} D_{\Gamma}(F,W) 
=- 2 D_{\Gamma}(F,W) \int_{l_2>0} P_{\Gamma}(\Gamma,l_2)  \\
&& + \int_{F_1> \Gamma} \int_{F_3 > \Gamma}
 P_{\Gamma}(F_1)
P_{\Gamma}(\Gamma) D_{\Gamma}(F_3,W)  \delta[F-(F_1+F_3-\Gamma) ]  \nonumber  \\
&& + \int_{F_1> \Gamma} \int_{F_3 > \Gamma}
 \int_{0< W'< F_1}
D_{\Gamma}(F_1,W')
P_{\Gamma}(\Gamma) P_{\Gamma}(F_3)  \delta[F-(F_1+F_3-\Gamma) ] 
\delta[W-(W'+F_3-\Gamma)]  \nonumber \\
&&+ \int_{F_1 > \Gamma} \int_{F_3 > \Gamma}
 \int_{0 < W' < \Gamma}    
 P_{\Gamma}(F_1) D_{\Gamma}(\Gamma,W')
 P_{\Gamma}(F_3)  \delta[F-(F_1+F_3-\Gamma) ] 
 \delta[W-(W'+F_3-\Gamma)] \nonumber
\end{eqnarray}

In the rescaled variables $\eta=\frac{F-\Gamma}{\Gamma}$ 
and $w=\frac{W}{\Gamma}$, the probability distribution
${\cal D}_{\Gamma}(\eta,w)=\Gamma^2 D_{\Gamma}(F,W)$ 
flows at large $\Gamma$
towards the fixed point ${\cal D}_*(\eta,w)$ satisfying
the equation
\begin{eqnarray} 
&& 0=(1+\eta) \partial_{\eta} {\cal D}^*(\eta,w)
+w \partial_w {\cal D}^*(\eta,w)
+\int_{max(0,w-1)}^{\eta} d\eta' e^{-(\eta-\eta') } {\cal D}^*(\eta',w) \\
&&+\int_0^{min(\eta,w)} d\eta' e^{-\eta'} {\cal D}^*(\eta-\eta',w-\eta')
+e^{-\eta} \int_{max(0,w-1)}^{min(\eta,w)} d\eta' {\cal D}^*(0,w-\eta')
\end{eqnarray}
in the domain $\eta \in [0,+\infty)$ and $0 \leq w \leq 1+\eta$.
Note that the probability that the origin belongs at scale
$\Gamma$ to a bond of barrier $\eta$ is already known (\ref{solu-sym})
\begin{eqnarray} \label{normalisationD}
\int_0^{1+\eta} dw  {\cal D}^*(\eta,w)
=\frac{1}{\overline{\lambda}} \int_0^{\infty} d\lambda \lambda
P^*(\eta,\lambda)=\frac{1+2 \eta}{3} e^{-\eta}
\end{eqnarray}
It is now more convenient to introduce the variable $v=1+\eta-w$
instead to $\eta$, and to set
${\cal D}^*(\eta,w)=e^{-\eta} \Phi(v=1+\eta-w,w)$.
The new function $\Phi(v,w)$ is now symmetric in $(v,w)$ and satisfies
\begin{eqnarray} \label{eqgenePhi}
&& 0=v \partial_v \Phi(v,w) +w \partial_w \Phi(v,w) -(v+w) \Phi(v,w)  \\
&&
+\int_{max(0,1-w)}^{v} dv' \Phi(v',w)+ \int_{max(0,1- v)}^{w} dw' \Phi(v,w')
+\int_{max(0,1-w)}^{min(1, v)} dv' \Phi(v',1-v')
\end{eqnarray}
on the domain $\{v \geq 0, w \geq 0, v+w \geq 1\}$.
The function  $J(v,w)=\partial_v \partial_w \Phi(v,w)$
satisfies the simpler equation
\begin{eqnarray} 
 0=\left( v \partial_v  +w \partial_w  +2-(v+w) \right) J(v,w)
\end{eqnarray}
and thus takes the form
\begin{eqnarray} 
 J(v,w) = C( \frac{v}{w}) \frac{e^{v+w}}{v w}
\end{eqnarray}
where $C(x)$ is an arbitrary function satisfying $C(x)=C(\frac{1}{x})$
to respect the symmetry $(v,w)$.
Since we are looking for a solution that is not exponentially
growing at $(v,w) \to \infty$, it is necessay to have $C(x) \equiv 0$.
The function $\Phi(v,w)$ is thus separable, and taking 
into account the symmetry in $(v,w)$, we have
\begin{eqnarray} \label{separable}
 \Phi(v,w)=\phi( v)+\phi(w) \\
\end{eqnarray}
where the function $\phi( v)$ has to satisfy the equation obtained
by replacing the form (\ref{separable}) into (\ref{eqgenePhi})
\begin{eqnarray} 
 0=v \phi'( v) - \left(v +max(0,1- v) \right) \phi( v)
+\int_{max(0,1- v)}^v dv' \phi(v') +\int_{max(0,1- v)}^{min(1, v)} dv' \phi(v') 
\end{eqnarray}
More explicitly, the equation becomes 
\begin{eqnarray} 
&& 0=v \phi'( v) - \phi( v)
+ 2 \int_{1-v}^v dv' \phi(v')  \ \ \ \ \   \hbox{for}  \ \  0 \leq v \leq 1  \\
&&  0=v \phi'( v) - v \phi( v)
+\int_0^v dv' \phi(v') +\int_{0}^{1} dv' \phi(v')
\ \ \ \ \  \hbox{for}  \ \   v \geq 1 
\end{eqnarray}

It is more convenient to derive these equations with respect to $v$
\begin{eqnarray} 
&&  0=v \phi''( v) 
+ 2  \phi( v) + 2  \phi(1- v) \ \ 
\ \ \  \hbox{for}  \ \  0 \leq v \leq 1 \\
&&  0=v \phi''( v) +(1-  v) \phi'( v)
\ \ \ \ \  \hbox{for}  \ \   v \geq 1 
\end{eqnarray}
and to keep the boundary condition at $v=1$
\begin{eqnarray} 
0=\phi'(1)-\phi(1)+2 \int_{0}^{1} dv' \phi(v')
\end{eqnarray}
In the domain $v \geq 1$,
the only solution that is not exponentially growing at $v \to \infty$
is the constant solution
\begin{eqnarray} 
\phi( v)=\phi(1) \ \  \hbox{for}  \ \   v \geq 1
\end{eqnarray}
and the condition at $v=1$ becomes
\begin{eqnarray} 
\phi(1)=2 \int_{0}^{1} dv' \phi(v')
\end{eqnarray}

The constant $\phi(1)$ has to be determined through 
the normalization condition (\ref{normalisationD})
\begin{eqnarray} 
\frac{1+2 \eta}{3}=\int_0^{1+\eta} dw \Phi(v=1+\eta-w,w)
= 2 \int_0^{1+\eta} dw \phi(w)= 2 \int_0^{1} dw \phi(w)
+2 \phi(1) \eta 
\end{eqnarray}
which leads to 
\begin{eqnarray} 
&& \phi(1)=\frac{1}{3} \\
&& \int_0^{1} dv \phi(v) =\frac{1}{6}
 \end{eqnarray}

We know consider the equation in the domain $v \in [0,1]$.
Since the equation is not local for $\phi( v)$, it is convenient
to introduce the sum
\begin{eqnarray} 
S( v)=\phi( v)+\phi(1- v)
 \end{eqnarray}
that satisfies the simpler equation
\begin{eqnarray} 
v (1- v) S''( v)+2 S( v)=0
 \end{eqnarray}
Two linearly independant solutions of this equation are
\begin{eqnarray} 
&& S^{\rm sym}( v)=v (1- v)  \\
&& S^{\rm antisym}( v)= 2 v (1- v) \ln \frac{v}{1-v} +2 v -1
 \end{eqnarray}
Since $S( v)$ has to be symmetric in $(v,1- v)$ by definition,
and has to be normalized to
\begin{eqnarray} 
\int_0^1 dv S( v) = 2 \int_0^1 dv \phi( v) = \frac{1}{3}
 \end{eqnarray}
we get
\begin{eqnarray} 
 S( v) = 2 v (1- v)
 \end{eqnarray}
and this corresponds to
\begin{eqnarray} 
 \phi( v)=\frac{1}{3}-\frac{2}{3}(1- v)^3
\end{eqnarray}
We thus finally get
\begin{eqnarray} 
&& \phi( v)= \frac{1}{3}-\frac{2}{3}(1- v)^3
\ \ \ \ \  \hbox{for}  \ \  0 \leq v \leq 1 \\
&& \phi( v)=\frac{1}{3} \ \ \ \ \  \hbox{for}  \ \   v \geq 1
\label{phi}
\end{eqnarray}

The final result of this section is that, at large $\Gamma$,
the probability ${\cal D}_*(\eta,w)$
that the point $x=0$ belongs 
to a bond of rescaled barrier $\eta=\frac{F-\Gamma}{\Gamma}$
 such that the point $x=0$ has a rescaled potential
$w=\frac{U(0)-U_{min}}{\Gamma}$ above the lower potential of the bond,
reads
\begin{eqnarray} 
 {\cal D}_*(\eta,w)= e^{-\eta} \Phi(1+\eta-w,w)
 = e^{-\eta} \left( \phi(w)+\phi(1+\eta-w) \right)
\end{eqnarray}

\begin{figure}[thb]

\centerline{\fig{12cm}{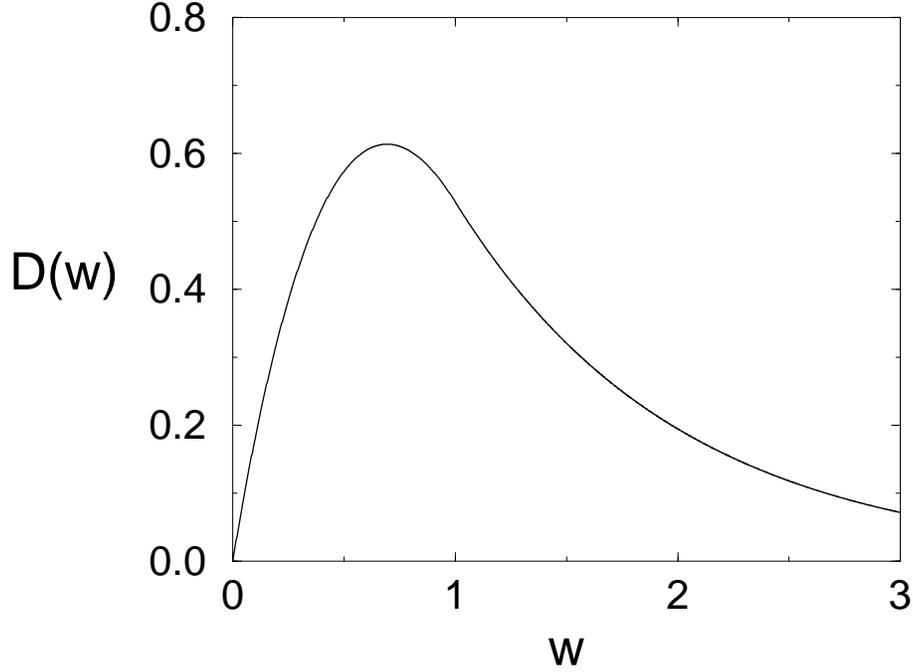}} 
\caption{ Exact asymptotic probability distribution $D_*(w)$
of the rescaled energy $w$. \label{fig4} } 

\end{figure}

In particular, the distribution of the rescaled energy $w$ (\ref{defw}) alone reads
\begin{eqnarray} 
&& {\cal D}_*(w)= \int_{max(0,w-1)}^{\infty} d\eta {\cal D}_*(\eta,w) \\
&&=  \theta(w \leq 1)  \left(4- 2 w-4 e^{-w} \right)
 +  \theta(w \geq 1)  \left( 2 e - 4 \right) e^{-w}
\label{res-sym}
\end{eqnarray}
This function is represented on Figure (\ref{fig4}).
It is continuous at $w=1$, as well as its first derivative,
but the second derivative is discontinuous at $w=1$.
The contribution to the normalization of the domain $w>1$ reads
\begin{eqnarray} 
\int_1^{+\infty} dw {\cal D}_*(w)=   2  - \frac{ 4 }{e} = 0.52848
\end{eqnarray}
The mean value is simply
\begin{eqnarray} 
\lim_{t \to \infty} \overline{ \frac{W(t)} {T \ln t} } =\int_0^{+\infty} dw w {\cal D}_*(w)=  \frac{4}{3}
\end{eqnarray}

\subsection{Distribution of energy for the biased case}

\label{energybias}

As discussed in details in Ref. \cite{us_sinai}, the RSRG method
can be applied to the Sinai diffusion in the presence of
a small external bias $\delta$, and the results are valid in the double
limit $\delta \to 0$ and $\Gamma \to \infty$ with the scaling variable
\begin{eqnarray} 
\gamma= \delta \Gamma = \delta T \ln t
\label{defgamma}
\end{eqnarray}
being fixed.

The above study of the symmetric case 
is generalized to the biased case in Appendix \ref{appbias}.
The final result for the
probability distribution $D_t(W)$ of $W(t)$ is
the sum of the contributions of the two types of bonds:
\begin{eqnarray} 
&& D_t(W) = D^+_\Gamma(W) + D^-_\Gamma(W) 
\end{eqnarray}
with $\Gamma=T \ln t$ and
\begin{eqnarray} 
 D_{\Gamma}^{\pm}(W>\Gamma) 
 && = \frac{\delta^2}{\sinh^2 \delta \Gamma
 ( \delta \coth \delta \Gamma \mp 3 \delta  ) }
\left( 1-2 e^{ - \Gamma ( \mp \delta +\delta  \coth \delta \Gamma)} \right)
e^{- (W-\Gamma) ( \mp \delta +\delta  \coth \delta \Gamma)}
\label{rgbiaswg1}
\end{eqnarray}
for $W>\Gamma$, and 
\begin{eqnarray} 
 D_{\Gamma}^{\pm}(W<\Gamma) 
 =  \frac{  \delta}{ 2 \sinh^2 \delta \Gamma  (\coth \delta \Gamma \mp 3  ) }
\left(- 4 e^{-W ( \mp \delta +\delta  \coth \delta \Gamma) }
+3 \mp \coth \delta \Gamma + ( 1 \pm  \coth \delta \Gamma) e^{ \mp 2 \delta W} \right)
\label{rgbiasws1}
\end{eqnarray}
for $0 \leq W \leq \Gamma$.

As in the symmetric case,
the probability distribution $D_{\Gamma}^{\pm}(W)$ and its first derivative
$\partial_W D_{\Gamma}^{\pm}(W)$ are continuous at $W=\Gamma$, but its second derivative $\partial_W^2 D_{\Gamma}^{\pm}(W)$ is discontinuous at $W=\Gamma$.

The full normalizations 
\begin{eqnarray} 
\int_0^{+\infty} dW D_{\Gamma}^{\pm}(W)
= \frac{ e^{ \pm 2 \gamma}-1 \mp 2\gamma}{4 \sinh^2 \gamma}
\end{eqnarray}
correspond as it should to
the ratios $\overline{l_{\Gamma}^{\pm}}/(\overline{l_{\Gamma}^{+}}+\overline{l_{\Gamma}^{-}})$
in terms of the mean lengths of the renormalized bonds. 

The mean value for the $+$ case reads
\begin{eqnarray} 
\overline{W_+(t)} && = \int_0^{+\infty} dW W D_{\Gamma}^{\pm}(W) \\
&& = 
\frac{ e^{ \gamma} (1-2 e^{ \gamma-\gamma \coth \gamma} )(e^{ \gamma}+\frac{\gamma}{\sinh\gamma})}{ \delta( \coth \gamma - 3 )}
+ 
\frac{ ( \frac{ (2 e^{6\gamma}-6 e^{4\gamma}+(5-2 \gamma^2) e^{2\gamma}-1+2\gamma+4\gamma^2)}{\sinh^2\gamma}-16(\gamma+e^{\gamma}\sinh \gamma ) e^{\gamma(3-\coth \gamma)})}{8 \delta (e^{2\gamma}-2)}
\end{eqnarray}
and varies between $\overline{W_+(t)} \to (2/3) \Gamma$ as $\delta \to 0$
and 
\begin{eqnarray} 
\overline{W_+(t)} && \opsimeq_{\gamma \to \infty} 
\frac{ 3 }{ 2 \delta} e^{ 2 \gamma}
\end{eqnarray}

\section{Joint distribution of energy and position}

\label{pathjoint}

In this section, we compute the joint distribution of position $x=x(t)-x(0)$
and energy $W=U(x(0))-U(x(t))$ by the direct path-integral approach
introduced in \cite{us_toy}.

\subsection{ Recall of the path-integral approach}

\label{sec-probalandscape}

As explained in details in \cite{us_toy}, the properties of the renormalized landscape can either be obtained as solutions of RSRG equations
or by the computation of constrained path-integrals.
In this second approach, the basic path integral  
\begin{eqnarray}
F_{[V_a,V_b]}(V_0, x_0, V , x )= 
\int_{V_{x_0}=V_0}^{V_{l}=V} DV(y) 
exp( - \int_0^x dy ( \frac{1}{4} (\frac{dV}{dy})^2  )
\theta_{[V_a,V_b]}^-(\{V\})
\label{fpath}
\end{eqnarray}
over the Brownian paths going from
$V(x_0)=V_0$ to $V(x)=V$ without having touched
the boundaries $V=V_a$ and $V=V_b$ can be expressed in Laplace transform 
with respect to $(x-x_0)$ as
\begin{eqnarray}
\hat{F}_{[V_a,V_b]}(V,p|V_0) = \frac{1}{W(p ; V_a,V_b)} 
\Phi_-(V,p; V_a) \Phi_+(V_0,p ; V_b) 
\qquad \text{if} ~~ V_a \leq V \leq V_0 \\
\hat{F}_{[V_a,V_b]}(V,p|V_0) = \frac{1}{W(p ; V_a,V_b)} 
\Phi_+(V,p ; V_b) \Phi_-(V_0,p ; V_a)
 \qquad \text{if} ~~ V_0 \leq V \leq V_b 
\label{res-system}
\end{eqnarray}
where $\Phi_-$ and $\Phi_+$ are solutions of the differential equation
\begin{eqnarray}
\partial^2_V \Phi  = p F
\end{eqnarray}
that vanish respectivly at $V=V_a$ and at $V=V_b$
\begin{eqnarray}
 \Phi_-(V,p;V_a) && = \sinh {\sqrt p} (V-V_a) \\
 \Phi_+(V,p;V_b) && =  \sinh {\sqrt p} (V_b- V) \\
\end{eqnarray}
and where $W(p ; V_a,V_b)$ is their wronskian
\begin{eqnarray}
 W(p ; V_a,V_b) && = {\sqrt p} \sinh {\sqrt p} (V_b-V_a)
\end{eqnarray}

As explained in \cite{us_toy}, the probability distribution $P_{\Gamma} ( F,l )$
corresponds to the path integral 
\begin{eqnarray}
B_{\Gamma} ( F,l ) = 
\int_{V_{0}=0}^{V_{l}=F} DV 
exp( - \int dx ( \frac{1}{4} (\frac{dV}{dx})^2  )
\Theta_\Gamma^-(\{V\})
\label{bpath}
\end{eqnarray}
where $F \ge  \Gamma$ and where
$\Theta_\Gamma^-(\{V\})$ constraints the paths to
remain in the interval $]0,F[$ for $0<x<l$ and
to not perform any returns of more than $\Gamma$,
i.e. for two arbitrary $0<x_1<x_2<l$, the potential has to satisfy
$V(x_1)-V(x_2) < \Gamma$. 
As shown in \cite{us_toy}, it can be obtained from the path-integral
(\ref{res-system}) as
\begin{eqnarray} 
{\hat B}_{\Gamma}(F,p )
  = \frac{ {\sqrt p} }{  \sinh {\sqrt p} \Gamma }
 e^{- (F-\Gamma) {\sqrt p}  \coth {\sqrt p} \Gamma}
\label{resB-sym}
\end{eqnarray}
in agreement with the fixed-point solution (\ref{solu-sym}) of the RSRG equations.

Similarly,
the probability $E_{\Gamma} (V,l,V_0)$
to go from $(V_0,0)$ to $(V,l)$ without making any return
of more than $\Gamma$ and where $V \geq V_0$ is the maximum
reads in Laplace \cite{us_toy}
\begin{eqnarray} 
{\hat E}_{\Gamma}(V,p,V_0 )
&&  =  e^{- (V-V_0) {\sqrt p}  \coth {\sqrt p} \Gamma}
\label{resE}
\end{eqnarray}

We are now in position to compute the joint distribution of energy and position
via a path-integral decomposition.

\subsection{Joint distribution of energy and position
for the symmetric case}

The law is symmetric in $x \to -x$ so we will consider $x>0$
and write the path-integral representation
\begin{eqnarray} 
 P_t(x>0, W)  && = \frac{1}{2 \overline{l_\Gamma}}
< \delta(V(x)- W)>_{\{V(y)\}} \\
&& =  \frac{1}{\Gamma^2}  
\int_v^{+\infty} dF \int_x^{+\infty} dl 
 \int_{V(0)=0}^{V(l)=F}
{\cal D }V(y) e^{-\frac{1}{4} \int_0^{l} dy 
\left( \frac{d V} {dy} \right)^2 }
\delta(V(x)- W)
\Theta_{\Gamma}^- \{V\}  
\label{jointpath} 
 \end{eqnarray}
where again $\Theta_{\Gamma} \{V(y)\}$ means that the path is constrained to remain in $(0,F)$ with no descending segment of more than $\Gamma$.
Since this constraint doesn't factorize into the two-sides $y<x$ and
$y>x$, we have to keep track of the maximum  
$V_m$ reached by $V(y)$ for $0<y<x$ to impose that the path doesn't 
go below $V_m-\Gamma$ for $x<y<l$.

We now compute the Laplace transform 
\begin{eqnarray} 
{\hat P}_t (p, W) && =\int_0^{+\infty} dx e^{-px}  P_t(x, W)   \\
\end{eqnarray}
for the two cases $W>\Gamma$ and $0<W<\Gamma$.

\subsubsection{ Case $W>\Gamma$  }

\begin{figure}[thb]

\centerline{\fig{12cm}{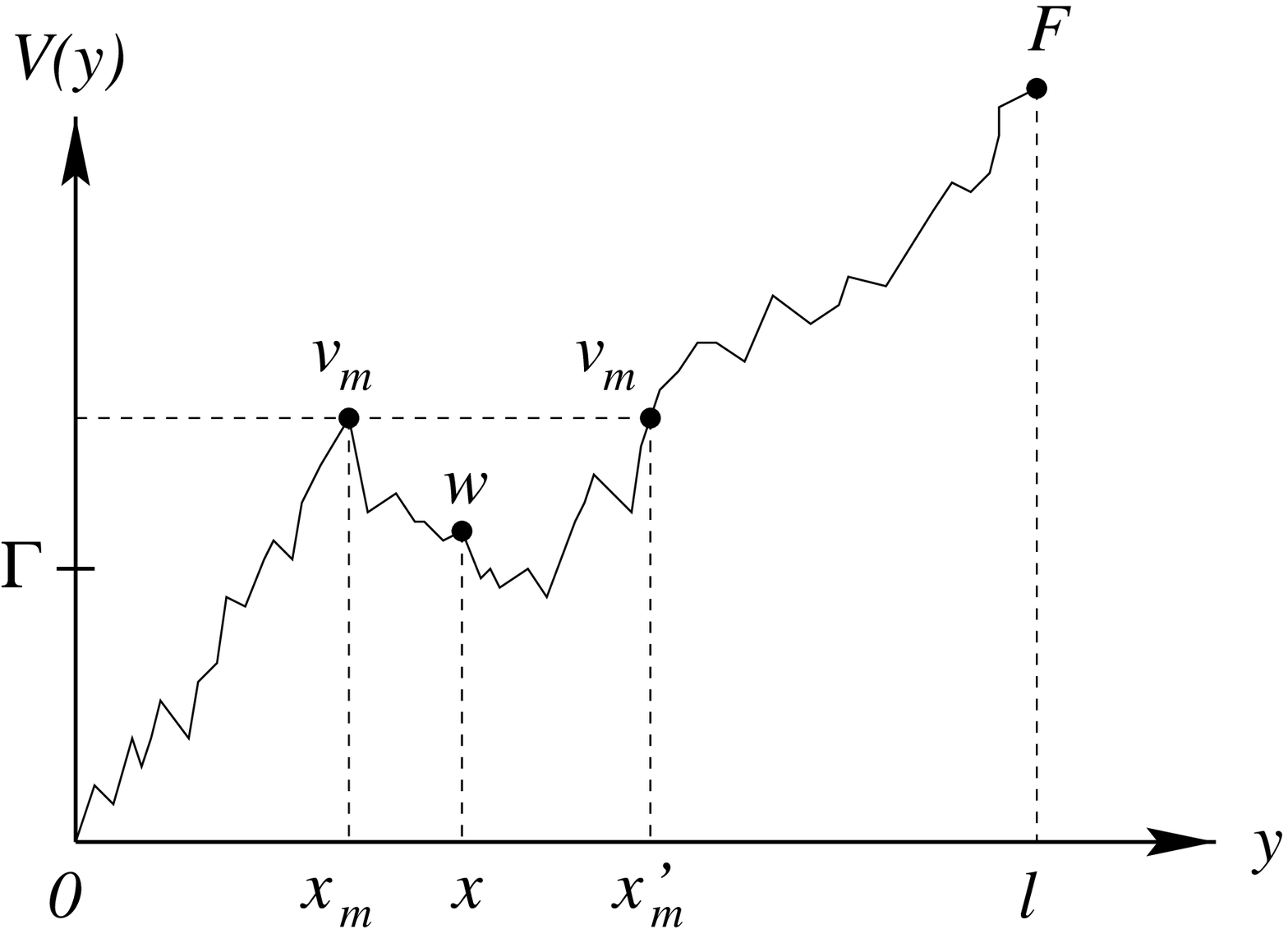}} 
\caption{ Properties of the renormalized bond for $W>\Gamma$. \label{fig1} } 

\end{figure}

Let us introduce the following notations (see Figure \ref{fig1}) :
We note $V_m$ the maximum reached by $ V(y)$ for $0<y<x$,
and $x_m$ the point where it is reached $V(x_m)=V_m$.
We note $x_m'$ the first time in $x<y<l$
where $ V(y)$ reaches again $V_m$. 
For the given parameters $(V_m,x_m,x_m')$, 
we have to sum over the Brownian paths satisfying the following conditions :
(i) they go from $(0,0)$ to $(V_m,x_m)$ without 
making any return of more than $\Gamma$,
and $V_m$ is reached for the first time at $x_m$;
(ii) they go from $(V_m,x_m)$ to $(W,x)$ without touching 
the absorbing boundaries at $V_m-\Gamma$ and at $V_m+\epsilon$.
(iii) they go from $(W,x)$ to $(V_m,x_m')$ without touching 
the absorbing boundaries at $V_m-\Gamma$ and at $V_m+\epsilon$.
(iv) they go from $(V_m,x_m')$ to $(F,l)$ without making any return
of more than $\Gamma$ and $F$ is the maximum.

The summation over the Brownian paths satisfying all these constraints 
leads to
\begin{eqnarray} 
{\hat P}_t (x, W>\Gamma)&& =  \frac{1}{\Gamma^2}  
 \int_W^{W+\Gamma} dV_m \int_0^x dx_m B_{\Gamma} (V_m,x_m) 
\left( \partial_{V_0} F_{[V_m-\Gamma,V_m]}(W,x\vert V_0,x_m )
\right) \vert_{V_0=V_m}  \\
&& \int_x^{+\infty} dx_m' 
(\partial_{V_1} F_{[V_m-\Gamma,V_m]}(V_1 ,x_m'\vert W,x )\vert_{V_1=V_m}
\int_{V_m}^{+\infty} dF \int_{x_m'}^{+\infty} dl
E_{\Gamma} (F,l;  V_m,x_m') 
\label{decomposition}
\end{eqnarray}
where the functions $F_{[V_a,V_b]}(V,x\vert V_0,x_0 )$,$B_{\Gamma} (F,l)$ 
and $E_{\Gamma} (F,l;  V,x)$ have been defined in section (\ref{sec-probalandscape})
and are given by the explicit expressions (\ref{res-system}) (\ref{resB-sym}) (\ref{resE}) in Laplace transform.

As a consequence, in Laplace transform with respect to $x$, we get
\begin{eqnarray} 
{\hat P}_t (p, W>\Gamma) && =   \frac{1}{\Gamma^2}
\int_W^{W+\Gamma} dV_m \frac{\sqrt p}{ \sinh {\sqrt p} \Gamma}
e^{-(V_m-\Gamma) {\sqrt p} \coth {\sqrt p} \Gamma} 
\frac{\sinh {\sqrt p} (\Gamma-(V_m- W))}{\sinh {\sqrt p} \Gamma}
\\  &&  \frac{(\Gamma-(V_m- W))}{\Gamma}
\int_{V_m}^{+\infty} dF e^{- \frac{F-V_m}{\Gamma} } \\
&& = \frac{ \sinh {\sqrt p} \Gamma}{ \Gamma^2 \sqrt p}
e^{-(W-\Gamma) {\sqrt p} \coth {\sqrt p} \Gamma} 
\left( 1-  2 \cosh {\sqrt p} \Gamma
 e^{-\Gamma {\sqrt p} \coth {\sqrt p} \Gamma} \right)
\end{eqnarray}

Rescaling the variables
\begin{eqnarray} 
{\cal P}(X=\frac{x}{\Gamma^2},w=\frac{W}{\Gamma})=
\Gamma^3 P_t(x, W) 
\label{rescaling}
\end{eqnarray}
we get the Laplace transform of the exact asymptotic distribution
of the rescaled variables $X$ and $w$
\begin{eqnarray} 
{\hat {\cal  P}}(s,w>1) && \equiv \int_0^{+\infty} dX e^{-sX} {\cal P}(X,w>1)  \\
&& = \frac{ \sinh {\sqrt s} }{ \sqrt s}
\left( e^{ {\sqrt s} \coth {\sqrt s} }-  2 \cosh {\sqrt s}   
\right)  e^{- w {\sqrt s} \coth {\sqrt s} } 
\label{reswb1}
\end{eqnarray}

In particular, the distribution of rescaled energy $w$ alone reads
(by taking into account the two sides $X>0$ and $X<0$)
\begin{eqnarray} 
 {\cal  P}(w>1) && = 2 {\hat {\cal  P}}(s=0,w>1)= 2 (e - 2) e^{-w}
\end{eqnarray}
in agrement with the RSRG result (\ref{res-sym}).

\subsubsection{ Case $W<\Gamma$  }

For $W<\Gamma$, there are two contributions coming from the cases
 $V_m>\Gamma$ and $V_m<\Gamma$, where $V_m$ is again the maximum
reached by the potential $V(y)$ for $0<y<x$
\begin{eqnarray} 
P_t (x, W<\Gamma) = P_t^{(1)} (x, W<\Gamma) + P_t^{(2)} (x, W<\Gamma)
\label{casemoins}
\end{eqnarray}

For $V_m>\Gamma$, we have the same decomposition as in (\ref{decomposition}),
except that here we have to integrate over $V_m \in (\Gamma,W+\Gamma)$
(instead of $V_m \in (W,W+\Gamma)$)
\begin{eqnarray} 
P_t^{(1)} (x, W)&& =  \frac{1}{\Gamma^2}  
 \int_{\Gamma}^{W+\Gamma} dV_m \int_0^x dx_m B_{\Gamma} (V_m,x_m) 
\left( \partial_{V_0} F_{[V_m-\Gamma,V_m]}(W,x\vert V_0,x_m )
\right) \vert_{V_0=V_m}  \\
&& \int_x^{+\infty} dx_m' 
(\partial_{V_1} F_{[V_m-\Gamma,V_m]}(V_1 ,x_m'\vert W,x )\vert_{V_1=V_m}
\int_{V_m}^{+\infty} dF \int_{x_m'}^{+\infty} dl
E_{\Gamma} (F,l;  V_m,x_m') 
\end{eqnarray}

In Laplace transform, we thus get
\begin{eqnarray} 
P^{(1)}(p, W<\Gamma)&& = \frac{1}{\Gamma^2} \int_{\Gamma}^{W+\Gamma}
 dV_m \frac{\sqrt p}{ \sinh {\sqrt p} \Gamma}
e^{-(V_m-\Gamma) {\sqrt p} \coth {\sqrt p} \Gamma} 
\frac{\sinh {\sqrt p} (\Gamma-(V_m- W))}{\sinh {\sqrt p} \Gamma}
\\ &&  \frac{(\Gamma-(V_m- W))}{\Gamma}
\int_{V_m}^{+\infty} dF e^{- \frac{F-V_m}{\Gamma} } \\
&& = \frac{1}{\Gamma^2} 
\left(  \frac{ \sinh {\sqrt p} (2 \Gamma- W)}{  {\sqrt p} }
- W  \frac{ \sinh {\sqrt p} (\Gamma- W)}{ \sinh {\sqrt p} \Gamma }
- \frac{ \sinh 2 {\sqrt p} \Gamma}{ {\sqrt p} } e^{- W
{\sqrt p} \coth {\sqrt p} \Gamma} \right)
\end{eqnarray}

Rescaling the variables as in (\ref{rescaling}), the first contribution
in (\ref{casemoins}) reads 
\begin{eqnarray} 
{\hat {\cal  P}}^{(1)}(s,w<1) && = \int_0^{+\infty} dX e^{-sX} {\cal P}^{(1)}(X,w<1) \\
&&   =  \frac{ \sinh {\sqrt s} (2-w)}{ {\sqrt s} }
- w  \frac{ \sinh {\sqrt s} (1-w)}{ \sinh {\sqrt s} }
- \frac{ \sinh 2 {\sqrt s} }{ {\sqrt s} } e^{- w {\sqrt s} \coth {\sqrt s}} 
\label{contri1}
\end{eqnarray}

\begin{figure}[thb]

\centerline{\fig{12cm}{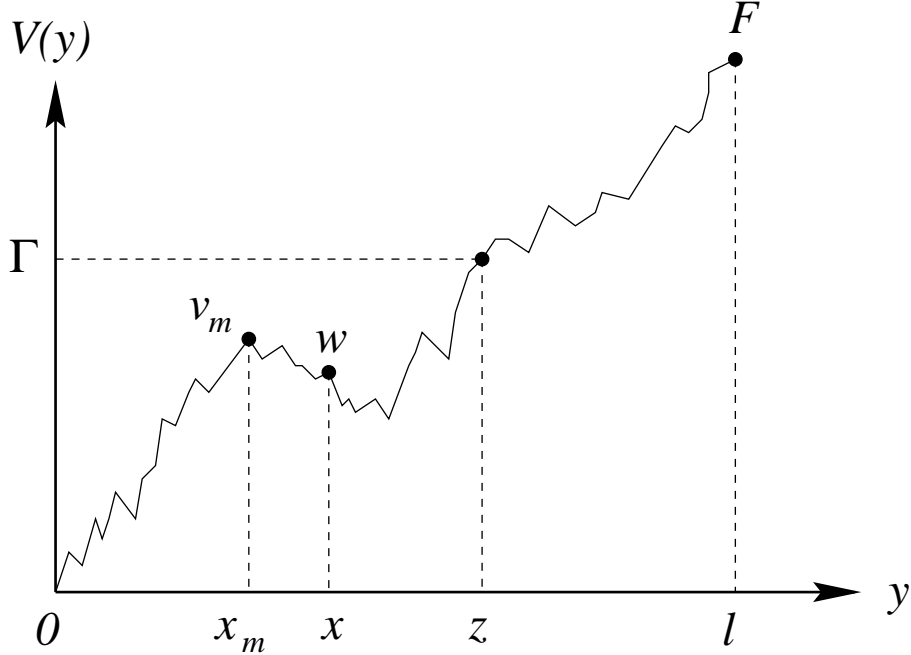}} 
\caption{ Properties of the renormalized bond for $V_m<\Gamma$. \label{fig2} } 

\end{figure}

We now consider the second contribution corresponding
to the cases where the maximum $V_m$ reached by $ V(y)$ for $0<y<x$
satisfies $V_m<\Gamma$ (see Figure \ref{fig2}).
We then have to sum over the Brownian paths satisfying
 the following conditions :
(i) they go from $(0,0)$ to $(W,x)$ without 
touching the bondaries at $V=0$ and $V=\Gamma$.
(ii) they go from $(W,x)$ to $(\Gamma,z)$ without touching 
the absorbing boundaries at $V=0$ and at $\gamma+\epsilon$.
(iv) they go from $(\Gamma,z)$ to $(F,l)$ without making any return
of more than $\Gamma$ and $F$ is the maximum.

The summation over the Brownian paths satisfying all these constraints 
leads to
\begin{eqnarray} 
{\hat P}_t^{(2)} (x, W) && =  \frac{1}{\Gamma^2} 
\left(\partial_{V_0} F_{[0,\Gamma]}(V_0,0;W,x) \right)_{V_0=0} 
\int_x^{\infty} dz \left(-\partial_{V'} F_{[0,\Gamma]}(W,x;,V',z) \right)_{V'=\Gamma} \\
&& \int_{\Gamma}^{+\infty} dF \int_z^{+\infty} dl E_{\Gamma}(\Gamma,z;F,l) 
\label{decompomoinssym}
\end{eqnarray}
and thus in Laplace with respect to $x$, we get

\begin{eqnarray} 
P^{(2)}(p, W)&& = \frac{1}{\Gamma^2} 
\left(\partial_{V_0} F_{[0,\Gamma]}(V_0,W;p) \right)_{V_0=0} 
\left(-\partial_{V'} F_{[0,\Gamma]}(W,V';0) \right)_{V'=\Gamma}
\int_{\Gamma}^{+\infty} dF E_{\Gamma}(\Gamma,F;0)  \\
&&   = W \frac{\sinh (\Gamma- W) \sqrt p}{ \Gamma^2 \sinh \Gamma \sqrt p}
\end{eqnarray}

Rescaling the variables as in (\ref{rescaling}), 
the second contribution in (\ref{casemoins}) reads

\begin{eqnarray} 
{\hat {\cal  P}}^{(2)}(s,w<1) && \equiv \int_0^{+\infty} dX e^{-sX} {\cal P}^{(1)}(X,w<1)=  w \frac{\sinh (1-w) \sqrt s}{\sinh  \sqrt s}
\label{contri2}
\end{eqnarray}

Adding the two contributions (\ref{contri1}) and (\ref{contri2}), we 
finally get the exact asymptotic distribution of the joined rescaled
variables $(X,w)$ in the domain $w<1$

\begin{eqnarray} 
{\hat {\cal  P}}(s,w<1) && = {\hat {\cal  P}}^{(1)}(s,w<1)
+ {\hat {\cal  P}}^{(2)}(s,w<1)\\
&& =    \frac{ \sinh {\sqrt s} (2-w)}{ {\sqrt s} }
- \frac{ \sinh 2 {\sqrt s} }{ {\sqrt s} } e^{- w {\sqrt s} \coth {\sqrt s}}
\label{resws1}
 \end{eqnarray}

In particular, the distribution of rescaled energy alone $w$ reads
\begin{eqnarray} 
 {\cal  P}(w<1) && = 2 {\hat {\cal  P}}(s=0,w<1) \\
&& = 2 ( 2-w -2 e^{-w} )
\end{eqnarray}
in agrement with the RSRG solution (\ref{res-sym}).

\subsubsection{Kesten distribution for the rescaled position alone }

A check of the above result is that one recovers the Kesten distribution,
obtained in \cite{us_sinai} via the RSRG, upon integration over the energy.

For the domain $w>1$, the distribution of the rescaled position $X$
has for Laplace transform (\ref{contri1})
\begin{eqnarray} 
{\hat {\cal  P}}_{>}(s) && = \int_1^{+\infty} dw {\hat {\cal  P}}(s,w>1) \\
&& = \frac{ \sinh^2 {\sqrt s} }{  s  }
\left( \frac{1}{   \cosh {\sqrt s}} - 2  e^{- {\sqrt s} \coth {\sqrt s} } 
\right)
\end{eqnarray}
whereas it is given in the domain $w<1$ by (\ref{contri2})
\begin{eqnarray} 
{\hat {\cal  P}}_{<}(s) && = \int_0^1 dw  {\hat {\cal  P}}(s,w<1) \\
&& = \frac{1}{s} \left( 1 -  \cosh {\sqrt s}
+ 2 \sinh^2 {\sqrt s} e^{- {\sqrt s} \coth {\sqrt s}}  \right)
\end{eqnarray}

Adding these two contributions, we recover as it should
the Kesten distribution \cite{us_sinai}
\begin{eqnarray} 
{\hat {\cal  P}}(s)  = {\hat {\cal  P}}_{<}(s) 
+ {\hat {\cal  P}}_{>}(s)  
 = \frac{1}{s} \left( 1- \frac{1} { \cosh {\sqrt s} }\right)
\end{eqnarray}

\subsubsection{Displacements for fixed energy}

It is interesting to use the above solution to compute the
moments of the distance traveled by the particle given that
the (rescaled) potential energy lost is a fixed $w=W/T \ln t$.
The probability distribution of the rescaled distance $X$ for a given rescaled energy $q_w(X)= {\cal P}(X,w) /\int dX {\cal P}(X,w)$
has for Laplace transform (\ref{reswb1},\ref{resws1}) 
\begin{eqnarray} \label{rescalq} 
q_w(s) && = \frac{{\hat {\cal  P}}(s,w>1)}{{\hat {\cal  P}}(0,w>1)} 
= \frac{ \sinh {\sqrt s} }{ \sqrt s (e-2)}
\left( e^{ {\sqrt s} \coth {\sqrt s} }-  2 \cosh {\sqrt s}   
\right)  e^{ w (1- {\sqrt s} \coth {\sqrt s} ) } 
\ \ \hbox{for} \ \ w>1
\\
q_w(s) && = \frac{{\hat {\cal  P}}(s,w<1)}{{\hat {\cal  P}}(0,w<1)} 
=
\frac{   \frac{ \sinh {\sqrt s} (2-w)}{ {\sqrt s} }
- \frac{ \sinh 2 {\sqrt s} }{ {\sqrt s} } e^{ - w {\sqrt s} \coth {\sqrt s}}  }{  (2-w)-2 e^{-w}}  \ \ \hbox{for} \ \ w<1
 \end{eqnarray}
The moments 
\begin{eqnarray} 
c_n(w) = \int_0^{+\infty} dX X^n q_w(X) =
< ( \frac{|x(t) - x(0)|}{(T \ln t)^2} ) ^n >_w  
\end{eqnarray} 
represent the moments of the distance traveled
 over the set of trajectories and
environments such that the potential energy lost is $w$.

In particular, the first moment reads
\begin{eqnarray} 
&& c_1(w) =  \frac{w}{3 } + \frac{8 - 3 e}{6 (e - 2)} \qquad 
 \ \ \hbox{for} \ \ w > 1 \\
&& c_1(w) = \frac{(2-w) (4 - e^w (2-w)^2)}{6(e^w (2-w) - 2)} 
 \ \ \hbox{for} \ \ \qquad w < 1
\end{eqnarray} 
and order $n$ similar polynomial extensions are obtained 
for the $n$-th moment. As a comparison, note that
$c_1$ is $5/12$ for the full Kesten diffusion front.

These moments $c_n(w)$ are increasing function of $w$ as expected since 
roughly speaking larger $w$ correspond to larger displacements.
For $w=0$, they simply vanish $c_n(w)=0$, which correspond to
\begin{eqnarray}  
q_{w=0}(s) =1   \ \ \hbox{i.e. } \ \ \qquad 
q_{w=0}(X)= \delta(X)     
 \end{eqnarray}
For large $w$, the asymptotic behavior of the moments is simply given by
\begin{eqnarray} 
c_n(w) \opsimeq_{w \to \infty} \left( \frac{w}{3} \right)^n
\end{eqnarray}
This interesting property 
 can be understood
by considering the large $w$ limit for the distribution
(\ref{rescalq}), which is dominated by the region $s \to 0$ :
\begin{eqnarray}  
q_w(s)   \opsimeq_{w \to \infty}   e^{ - w \frac{s}{3}  } 
 \end{eqnarray}
corresponding after Laplace inversion to
\begin{eqnarray}  
q_w(X)   \opsimeq_{w \to \infty}   \delta(X- \frac{w}{3} )  
 \end{eqnarray}
 i.e as $w$ becomes large, in terms of the original variables, we have
\begin{eqnarray} 
&& \frac{x(t)}{(T \ln t)^2}  = \frac{1}{3}  \frac{W(t)}{T \ln t}
\end{eqnarray}
with probability one.

\subsection{Joint distribution of energy and position
for the biased case}

The above calculation can be generalized in the presence of a small bias.

The joint distribution of energy and position is given for the biased case
by the following path-integrals with $x>0$
\begin{eqnarray} 
 P_t(\pm x, W)  && = \frac{1}{\overline{l_+}+\overline{l_-}}
< \delta(V_{\pm}(x)- W)>_{\{V_{\pm}(y)\}}  \\
&& = \frac{\delta^2}{\sinh^2 \delta \Gamma} 
\int_W^{+\infty} dF \int_x^{+\infty} dl 
 \int_{V(0)=0}^{V(l)=F}
{\cal D } V(y) e^{-\frac{1}{4} \int_0^{l} dy 
\left( \frac{d V} {dy} \mp 2 \delta\right)^2 }
\Theta_{\Gamma} \{V(y)\}  \delta(V(x)- W)
\label{jointpathbiased} 
\end{eqnarray}

The explicit computation is done in Appendix (\ref{jointbias})
and leads to the following final results in Laplace transform
for $W>\Gamma$ and $W<\Gamma$ respectivly
\begin{eqnarray} 
{\hat P}_t^{\pm} (p, W>\Gamma) && =  \frac{\delta^2 }
{\sinh^2 \delta \Gamma}
\frac{  \frac{\sqrt  {p +\delta^2}}{\sinh {\sqrt  {p +\delta^2}} \Gamma } 
e^{\mp \delta \Gamma}-2 (\mp \delta+\delta \coth \delta \Gamma ) 
e^{- \Gamma (\mp \delta+\delta \coth \delta \Gamma ) } }
{ ( \frac{\delta^2+p}{\sinh^2 {\sqrt  {p +\delta^2}} \Gamma }
+4 \delta^2 
\mp 4 \delta {\sqrt  {p +\delta^2}} \coth {\sqrt  {p +\delta^2}} \Gamma ) }  e^{- (W-\Gamma) ( \mp \delta +{\sqrt  {p +\delta^2}}  \coth {\sqrt  {p +\delta^2}} \Gamma)}
\label{biasjointplus}
\end{eqnarray}
\begin{eqnarray} 
&& {\hat P}_t^{\pm} (p, W<\Gamma) 
 = \frac{\delta^2}{\sinh^2 \delta \Gamma \left(
\frac{p+\delta^2}{\sinh^2 {\sqrt  {p +\delta^2}} \Gamma} \mp 4 \delta ( \mp \delta + {\sqrt  {p +\delta^2}}  \coth {\sqrt  {p +\delta^2}} \Gamma)  \right) } \\
&& [-2   ( \mp \delta +{\sqrt  {p +\delta^2}}  \coth {\sqrt  {p +\delta^2}} \Gamma)
 e^{-W ( \mp \delta +{\sqrt  {p +\delta^2}}  \coth {\sqrt  {p +\delta^2}} \Gamma)} \\
&& +2 ( \mp \delta +{\sqrt  {p +\delta^2}}  \coth {\sqrt  {p +\delta^2}} \Gamma) \cosh {\sqrt  {p +\delta^2}} W 
\cosh \delta W \\
&& -  ({\sqrt  {p +\delta^2}}(1+\coth^2 {\sqrt {p+\delta^2}} \Gamma)
\mp 2 \delta \coth {\sqrt  {p +\delta^2}} \Gamma) ) 
\sinh {\sqrt  {p +\delta^2}}  W \cosh \delta W \\
&& + \frac{1}{\delta} (p-( \mp \delta +{\sqrt  {p +\delta^2}}  \coth {\sqrt  {p +\delta^2}} \Gamma)^2)\cosh {\sqrt  {p +\delta^2}}  W \sinh \delta W  \\
&& +  ( \mp \delta +{\sqrt  {p +\delta^2}}  \coth {\sqrt  {p +\delta^2}} \Gamma )(-2  \coth {\sqrt  {p +\delta^2}} \Gamma +
\frac{{\sqrt {p+\delta^2}}}{ \delta \sinh^2 {\sqrt {p+\delta^2}} \Gamma})\sinh {\sqrt  {p +\delta^2}}  W \sinh \delta W] \\
&& + \frac{\delta \sinh \delta W \sinh (\Gamma- W) {\sqrt {p+\delta^2}} 
}{\sinh^2 \delta \Gamma \sinh \Gamma {\sqrt {p+\delta^2}}}
\label{biasjointmoins}
\end{eqnarray}

For the special case $p=0$, these expressions coincide as it should
with the results (\ref{rgbiaswg1}) and (\ref{rgbiasws1}) corresponding
to the distribution of the energy $W$ alone.  

On the other hand, the integration over the energy $W$
gives
\begin{eqnarray} 
{\hat P}_t^{>\pm} (p) && = \int_{\Gamma}^{+\infty} dW {\hat P}_t^{\pm} (p, W>\Gamma) \\
&& =   \frac{\delta^2 }
{\sinh^2 \delta \Gamma}
\frac{  \frac{\sqrt  {p +\delta^2}}{\sinh {\sqrt  {p +\delta^2}} \Gamma } 
e^{\mp \delta \Gamma}-2 (\mp \delta+\delta \coth \delta \Gamma ) 
e^{- \Gamma (\mp \delta+\delta \coth \delta \Gamma ) } }
{ ( \frac{\delta^2+p}{\sinh^2 {\sqrt  {p +\delta^2}} \Gamma }
+4 \delta^2 
\mp 4 \delta {\sqrt  {p +\delta^2}} \coth {\sqrt  {p +\delta^2}} \Gamma )
 ( \mp \delta +{\sqrt  {p +\delta^2}}  \coth {\sqrt  {p +\delta^2}} \Gamma) }  
\end{eqnarray}
and
\begin{eqnarray} 
{\hat P}_t^{<\pm} (p) && = \int_0^{+\Gamma} dW  {\hat P}_t^{\pm} (p, W<\Gamma) \\
&& = \frac{ \delta^2 } {p \sinh^2 \delta \Gamma
 (({\sqrt {p+\delta^2}} \coth {\sqrt {p+\delta^2}} \Gamma-2 \delta)^2-p-\delta^2))}
[2  p e^{ - \Gamma ( \mp \delta +{\sqrt  {p +\delta^2}}  \coth {\sqrt  {p +\delta^2}} \Gamma ) } \\ 
&& +
\frac{{\sqrt {p+\delta^2}} \sinh \delta }{ \delta \sinh {\sqrt {p+\delta^2}} \Gamma}(\delta^2+3 \delta^2 \coth \delta \Gamma-  \delta  {\sqrt {p+\delta^2}} \coth {\sqrt {p+\delta^2}}
 \Gamma (3+\coth \delta \Gamma)+ \frac{p+\delta^2}{\sinh^2 {\sqrt {p+\delta^2}} \Gamma})] 
 \\
&& +  \frac{\delta^2}{p \sinh^2 \delta \Gamma }
\left( 1
- \frac{{\sqrt {p+\delta^2}} \sinh \delta \Gamma }{ \delta \sinh  {\sqrt {p+\delta^2}} \Gamma }   \right)
\end{eqnarray}
whose sum coincides as it should with the distribution of the
position alone \cite{us_sinai}
\begin{eqnarray} 
{\hat P}_t^{\pm} (p) && =  {\hat P}_t^{>\pm} (p)+ {\hat P}_t^{<\pm} (p) \\
&& = \frac{ \delta^2}{p \sinh^2 \delta \Gamma}
 \left(1- \frac{ \frac{\sqrt {p+\delta^2}}{\sinh {\sqrt {p+\delta^2}} \Gamma} e^{\mp \delta \Gamma} }{  ( \pm \delta +{\sqrt  {p +\delta^2}}  \coth {\sqrt  {p +\delta^2}} \Gamma)  } \right)
\end{eqnarray}

\section{Results in the presence of a reflecting boundary}

\label{reflecting}

In this section, we consider the case where there is a reflecting boundary
at the starting point $x=0$ \cite{us_sinai}, with possibly some bias. It is
interesting as it can be studied easily using the RSRG. 

\subsection{Joint distribution of energy and position
on the half-line}

In the case of a bias, the joint probability distribution of
$E_{\Gamma}^{\pm}(W,l)$ of the energy-loss $W=U(t=0)-U(t)$
and the distance $l=x(t)-x(0)$
 satisfies the following RG equation

\begin{eqnarray} 
&& \partial_\Gamma  E_{\Gamma}^{\pm}(W,l) = 
  - E_{\Gamma}^{\pm}(W,l) \int_0^{\infty} dl' P_{\Gamma}^{\mp}(\Gamma,l') \\
&&
+ \int_0^{W} dW_1 \int_0^{\infty} dl_1 \int_0^{\infty} dl_2  \int_{\Gamma}^{\infty} dF_3 \int_0^{\infty} dl_3
E_{\Gamma}^{\pm} (W_1,l_1)  P_{\Gamma}^{\mp}(\Gamma,l_2)  P_{\Gamma}^{\pm}(F_3 ,l_3)
\\ && 
\delta(W-(W_1+F_3-\Gamma)) \delta(l-(l_1+l_2+l_3))
\label{brsrgdrift}
 \end{eqnarray}

For large $\Gamma$ using the properties of the fixed point
solution (\ref{solu-biased}) for $P^\pm$, and
the properties (\ref{equpm}) of the functions $U$ and $u$,
this can be rewritten in the Laplace variable with respect to the length as
\begin{eqnarray}
\partial_\Gamma  E_\Gamma^{\pm}(W,p) = 
 - u_\Gamma^{\mp}(0) E_{\Gamma}^{\pm}(W,p) 
+ U_\Gamma^{+}(p) U_\Gamma^{-}(p) \int_0^{W} dW_1 
E_{\Gamma}^{\pm} (W_1,p)  e^{-(W-W_1) u_\Gamma^{\pm}(p) }
\end{eqnarray}
and we finally get that the joint distribution
of position $l$ and energy $W$ takes the very simple form
\begin{eqnarray}
E_\Gamma^{\pm}(W,p) = u_\Gamma^{\pm}(0) e^{-W u_\Gamma^{\pm}(p)} 
\label{finitex}
\end{eqnarray}

The distribution of the energy $W$ alone is a simple exponential
\begin{eqnarray}
E_\Gamma^{\pm}(W) = u_\Gamma^{\pm}(0) e^{-W u_\Gamma^{\pm}(0)} 
\end{eqnarray}
contrary to the case studied in previous sections
where the diffusion takes place on the full line. The absence of
nonanalytic behaviour is traced to the fact that here decimations of
the bond near the boundary do not occur.

\subsection{Aging for the joint distribution on the half-line}

It is now possible to obtain the correlation of the energy at two
different times, i.e. the aging of the system in energy.

In the case of a reflecting boundary at the origin,
the single time diffusion front is simply 
\begin{eqnarray}
\overline{P(W',x',t'|0,0,0)} = E_{\Gamma'}^{+}(W',x')
\end{eqnarray}
whose Laplace transform is given in equation (\ref{finitex}).
Similarly, the two-time diffusion front 
$\overline{P(W,x,t,W',x',t'|00)} = E^{+}_{\Gamma,\Gamma'}(W,x,W',x')$,
satisfies the RG equation given in 
(\ref{brsrgdrift})
but with the initial condition  at $\Gamma=\Gamma'$ :
\begin{eqnarray}
E_{\Gamma=\Gamma',\Gamma'}^{+}(W,x,W',x') 
= \delta(W-W')\delta(x-x') E_{\Gamma'}^{+}(W',x')
\end{eqnarray}

Within the effective dynamics, this system
has the flavor of a directed model, since $W-W'$ and
$x(t)-x(t')$ are always positive. It is thus convenient to
define the Laplace transforms
\begin{eqnarray}
\hat{E}_{\Gamma,\Gamma'}^{+}(\lambda,p,W',p') 
=\int_0^{+\infty} dx' e^{-p'x'} \int_{x'}^{\infty} dx e^{-p(x-x')} \int_{W'}^{\infty} dW e^{-\lambda(W-W')} 
E_{\Gamma,\Gamma'}^{+}(W,x,W',x')
\end{eqnarray}

Using the fixed point solution (\ref{solu-biased})
for $P_{\Gamma}^{\pm}$ together with the properties (\ref{equpm}),
the above RG equation simplifies into 
\begin{eqnarray}
\partial_\Gamma  \ln \hat{E}_{\Gamma,\Gamma'}^{+}(\lambda,p,W',p') = 
 - u_\Gamma^{-}(0)  
+ \frac{ U_\Gamma^{+}(p) U_\Gamma^{-}(p) } {\lambda+ u_\Gamma^{+}(p) }
= \partial_\Gamma  \ln \left( \frac{u_\Gamma^{+}(0)}{\lambda+ u_\Gamma^{+}(p)} \right)
\end{eqnarray}
with the initial condition
\begin{eqnarray}
\hat{E}_{\Gamma=\Gamma',\Gamma'}^{+}(\lambda,p,W',p') 
=     E_{\Gamma'}^{+}(W',p') = u_{\Gamma'}^{+}(0) e^{-W' u_{\Gamma'}^{+}(p')}
\end{eqnarray}

We thus obtain
\begin{eqnarray}
\hat{E}_{\Gamma,\Gamma'}^{+}(\lambda,p,W',p') = E_{\Gamma'}^{+}(W',p')
 \frac{u_\Gamma^{+}(0)}{u_{\Gamma'}^{+}(0)} \ \ \frac{\lambda+ u_{\Gamma'}^{+}(p)} {\lambda+ u_\Gamma^{+}(p)} 
\end{eqnarray}

or more expliciltly in terms of $(W,W')$ and $(p,p')$
\begin{eqnarray}
\hat{E}_{\Gamma,\Gamma'}^{+}(W,p,W',p') = u_{\Gamma}^{+}(0) e^{-W' u_{\Gamma'}^{+}(p')}
\left( \delta(W-W') + 
 ( u_{\Gamma'}^{+}(p) - u_\Gamma^{+}(p) ) e^{-(W-W') u_\Gamma^{+}(p) }  \right)
\end{eqnarray}

In particular for $p=0=p'$, we obtain the aging properties of
the energy alone
\begin{eqnarray}
E_{\Gamma,\Gamma'}^{+}(W,W') = u_{\Gamma}^{+}(0) e^{-W' u_{\Gamma'}^{+}(0)}
\left( \delta(W-W') + 
 ( u_{\Gamma'}^{+}(0) - u_\Gamma^{+}(0) ) e^{-(W-W') u_\Gamma^{+}(0) }  \right)
\end{eqnarray}

For the symmetric case ($\delta=0$), it simplifies into

\begin{eqnarray}
E_{\Gamma,\Gamma'}(W,W') = \frac{1}{\Gamma} e^{-W'/ \Gamma'}
\left( \delta(W-W') + 
 ( \frac{1}{\Gamma'}-\frac{1}{\Gamma}) e^{-(W-W') /\Gamma}   \right)
\end{eqnarray}

In particular, the two-time Energy Correlation reads

\begin{eqnarray}
\overline{U(t) U(t')} = \int_0^{+\infty} dW' \int_{W'}^{+\infty} dW   \ W W' E_{\Gamma,\Gamma'}^{+}(W,W') = \frac{1}{u_{\Gamma'}^{+}(0)} 
\left( \frac{1}{u_{\Gamma'}^{+}(0)} + \frac{1}{u_{\Gamma}^{+}(0)} \right)
\end{eqnarray}

In the limit $\gamma= \Gamma \delta = T \delta \ln t \gg 1$
$\gamma'= \Gamma' \delta = T \delta \ln t' \gg 1$, 
the influence of the
wall vanishes in that limit, and the result should coincide with the result of the full model
without a wall.

\section{distribution of the magnetization in the 1D spin glass}

\label{sg}

Finally, let us indicate how the above results can be translated in the
language of the one dimensional Ising spin glass in an external field. We refer the reader to
\cite{us_rfim} for the details of the study of the spin glass by the
RSRG method. We consider the Hamiltonian 
\begin{eqnarray}
H = - \sum_{i=1}^{i=N-1} J_i \sigma_i \sigma_{i+1} - \sum_{i=1}^{i=N} h \sigma_i
\end{eqnarray}
with a bimodal distribution of random bonds $J_i = J \epsilon_i$. 
In the absence of a field $h=0$ the spin glass has two ground states
$\pm \sigma_i^0$ with $\sigma_i^0 = \epsilon_1 ..\epsilon_{i-1}$. In the presence
of a field $h$ the ground state of the spin glass is made of domains of either
zero field ground states. At finite time $t$ after the quench, these
domain grow until they reach their equilibrium size at equilibration
time $\Gamma_{eq} = T \ln t_{eq} = 4 J$. Each of these domain consists of
an interval between two frustrated bonds (i.e. such that
$J_i \sigma_i \sigma_{i+1} < 0$). During the coarsening process 
the number of such bonds decrease with time.

Choosing a point $O$ randomly in space one can thus, at time $t$, 
corresponding to the RSRG scale $\Gamma=T \ln t$
denote by $i=R$ (and $i=L$ respectively) the closest frustrated bond to the right (respectively the left) of point $0$. 
In the spin glass the (absolute value of the)
magnetization plays the role of the energy barriers in the Sinai model, more
precisely:
\begin{eqnarray}
|M_{L0}| \equiv | \sum_{L \leq i \leq O} \sigma_i |  =
| \sum_{L \leq i \leq O} \sigma_i^0 |  =
\frac{1}{2 h} |V(L) - V(0)|
\end{eqnarray}
where $V$ is a Sinai type potential. It is then easy to see that if one 
defines 
\begin{eqnarray}
w_1 = \frac{|M_{L0}(t)|}{T \ln t} \qquad w_2 = \frac{|M_{0R}(t)|}{T \ln t} 
\label{def}
\end{eqnarray}
then the joint distribution $P(w_1,w_2)$ of the scaling variables $w_1,w_2$ has been
determined in Section II:
\begin{eqnarray} 
P(w_1,w_2) =  {\cal D}_* (\eta = w_1 + w_2 - 1 , w_1) = e^{- w_1 - w_2 + 1} (\phi(w_1) + \phi(w_2))
\theta(w_1 + w_2 - 1)
\label{eqmag}
\end{eqnarray}
where $\phi(w)$ is given in (\ref{phi}). It exhibits, in particular, the same
nonanalyticity as the distribution of the potential energy loss
of a particle in the Sinai model. The distribution of the total
magnetization $|M_{L0}(t) + M_{0R}(t)| = |M_{L0}(t)| +   |M_{0R}(t)|$
is just the barrier distribution, i.e. a simple exponential,
as found in \cite{us_rfim}. Note also, that the joint distribution
of the {\it equilibrium} magnetizations is also (\ref{eqmag}) with
the definitions (\ref{def}) replacing $T \ln t \to 4 J$.

\appendix

\section{Generalization to the biased case}

\label{appbias}

\subsection{Recall of main results of the real space renormalization method}

In the presence of an external drift, there are
two probability distributions $P_{\Gamma}^{\pm}(F,l)$
describing respectivly the ascending and the descending bonds.
The asymptotic distributions read \cite{us_sinai}
\begin{eqnarray} 
\label{solu-biased}
P_{\Gamma}^{\pm}(F,p) & = & U_{\Gamma}^{\pm}(p) e^{- (F-\Gamma) u_{\Gamma}^{\pm}(p)} \\
u_{\Gamma}^{\pm}(p) & = & \sqrt{p + \delta^2} \coth{[\Gamma \sqrt{p + \delta^2}]} 
\mp \delta  \\
U_{\Gamma}^{\pm}(p)  & = & \frac{\sqrt{p 
+ \delta^2}}{\sinh{[\Gamma \sqrt{p + \delta^2}]}} e^{\mp \delta \Gamma}
\end{eqnarray}

It is also usefull to recall the evolution equations
\begin{eqnarray} 
\label{equpm}
&& \partial_\Gamma u^{\pm}_\Gamma (p) 
= - U^{+}_\Gamma (p) U^{-}_\Gamma (p) \\
&& \partial_\Gamma U^{\pm}_\Gamma (p) 
= - u^{\mp}_\Gamma (p) U^{\pm}_\Gamma (p) 
\end{eqnarray}

In particular, the distributions of the barriers 
alone are in terms of the variable $z=F-\Gamma$

\begin{eqnarray}
&&  P_{\Gamma}^{\pm} (z)
= u^{\pm}_{\Gamma}(0) e^{-z u^{\pm}_{\Gamma}(0)}   \\
&& u^{\pm}_{\Gamma}(0)=\frac{\delta}{\sinh (\Gamma \delta)}
e^{\mp \delta \Gamma}
\label{zdrift}
 \end{eqnarray}

Finally, it is also convenient to introduce the density of renormalized valleys
\begin{eqnarray}
n_{\Gamma} =\frac{1}{\overline{l^+_{\Gamma}}+\overline{l^-_{\Gamma}}}
= \frac{\delta^2}{\sinh^2 (\Gamma \delta)} = u^{+}_{\Gamma}(0) u^{-}_{\Gamma}(0)
\label{ngamma}
 \end{eqnarray}

\subsection{Distribution of energy via the RSRG equation}

We introduce the probability $D_{\Gamma}^{\pm}(F,W)$
that the point $x=0$ belongs at scale $\Gamma$
to a $\pm$ bond of barrier $F$ and
 is at potential $W$ above the lower potential of the bond
(i.e. $U(x=0)-U_{min}=W$).
They are normalized to
\begin{eqnarray} 
\int_0^{F} dW D_{\Gamma}^{\pm}(F,W)=
\frac{1}{\overline{l^+_{\Gamma}}+\overline{l^-_{\Gamma}}} 
\int_0^{\infty} dl l P_{\Gamma}^{\pm} (F,l)
\label{normadrift}
\end{eqnarray}

The RG equations for $D_{\Gamma}^{\pm}(F,W)$ are
\begin{eqnarray} 
&&  \partial_{\Gamma} D_{\Gamma}^{\pm}(F,W) 
=- 2 D_{\Gamma}^{\pm}(F,W)  P_{\Gamma}^{\mp}(\Gamma) \\
&& + \int_{\Gamma}^{\infty} dF_1 \int_{\Gamma}^{\infty} dF_3
 P_{\Gamma}^{\pm}(F_1)
P_{\Gamma}^{\mp}(\Gamma) D_{\Gamma}^{\pm}(F_3,W) 
 \delta[F-(F_1+F_3-\Gamma) ]   \\
&&+ \int_{\Gamma}^{\infty} dF_1 \int_{\Gamma}^{\infty} dF_3
 \int_0^{F_1} dW'
D_{\Gamma}^{\pm}(F_1,W')
P_{\Gamma}^{\mp}(\Gamma) P_{\Gamma}^{\pm}(F_3) 
 \delta[F-(F_1+F_3-\Gamma) ] 
\delta[W-(W'+F_3-\Gamma)]   \\
&&+ \int_{\Gamma}^{\infty} dF_1 \int_{\Gamma}^{\infty} dF_3
 \int_0^{\Gamma} dW'    
 P_{\Gamma}^{\pm}(F_1) E_{\Gamma}^{\mp}(\Gamma,W')
 P_{\Gamma}^{\pm}(F_3)  \delta[F-(F_1+F_3-\Gamma) ] 
 \delta[W-(W'+F_3-\Gamma)] 
\end{eqnarray}

Using the variable $z=F-\Gamma$ and the fixed point solutions 
$ P_{\Gamma}^{\pm} (z)$ (\ref{zdrift}), the RG equations become

\begin{eqnarray} 
&&  (\partial_{\Gamma} - \partial_z) D_{\Gamma}^{\pm}(z,W) 
=- 2 u^{\mp}_{\Gamma}  D_{\Gamma}^{\pm}(z,W)  + u^{-}_{\Gamma} u^{+}_{\Gamma} 
\int_{max(0,W-\Gamma)}^{z} dz' e^{-(z-z') u^{\pm}_{\Gamma}}
D_{\Gamma}^{\pm}(z',W) \\
&& + u^{-}_{\Gamma} u^{+}_{\Gamma} 
\int_{0}^{min(z,W)} dz' e^{-z' u^{\pm}_{\Gamma}}
D_{\Gamma}^{\pm}(z-z',W-z') + (u^{\pm}_{\Gamma})^2 e^{-z u^{\pm}_{\Gamma}}
\int_{max(0,W-\Gamma)}^{min(z,W)} dz' D_{\Gamma}^{\mp}(0,W-z') 
\end{eqnarray}

Using (\ref{ngamma}) it is convenient to set
\begin{eqnarray} 
 D_{\Gamma}^{\pm}(z,W) 
=n_{\Gamma} u^{\pm}_{\Gamma} e^{- z u^{\pm}_{\Gamma}}
\Psi^{\pm}_{\Gamma}(V=z+\Gamma-W,W) 
\end{eqnarray}
and to write the RG equations for $\Psi^{\pm}_{\Gamma}(V,W)$

\begin{eqnarray} 
&& \left[ \partial_{\Gamma}+(V+W-\Gamma) n_{\Gamma} \right] \Psi_{\Gamma}^{\pm}(V,W) \\ 
=&&  n_{\Gamma} \left( \int_{max(0,\Gamma-W)}^{V} dV' 
 \Psi_{\Gamma}^{\pm}(V',W)+ 
\int_{max(0,\Gamma-V)}^{W} dW'
 \Psi_{\Gamma}^{\pm}(V,W') +\int_{max(0,\Gamma-W)}^{min(\Gamma,V)} dV' 
 \Psi_{\Gamma}^{\mp}(V',\Gamma-V') \right)
\end{eqnarray}

Applying $\partial_V \partial_W $ to these equations lead to
the simple equations
\begin{eqnarray} 
 \left[ \partial_{\Gamma}+(V+W-\Gamma) n_{\Gamma} \right]
 \partial_V \partial_W \Psi_{\Gamma}^{\pm}(V,W)=0 
\end{eqnarray}
Since we are loking for solutions that are not exponentially growing as $(V,W)
\to +\infty$, we are led, as in the symmetric case, to

\begin{eqnarray} 
 \partial_V \partial_W \Psi_{\Gamma}^{\pm}(V,W)=0 
\end{eqnarray}

It is now natural to set
\begin{eqnarray} 
\Psi_{\Gamma}^{+}(V,W)=\Psi_{\Gamma}^{-}(V,W)=\Psi_{\Gamma}(V,W)
=\psi_{\Gamma}(V)+\psi_{\Gamma}(W)
\end{eqnarray}
and the equation for $\psi_{\Gamma}(V)$ reads

\begin{eqnarray} 
&& \left[ \partial_{\Gamma} +(V-\Gamma+max(0,\Gamma-V)) n_{\Gamma} \right] \psi_{\Gamma}(V) \\ 
=&& + n_{\Gamma} \left( \int_{max(0,\Gamma-V)}^{V} dV' 
 \psi_{\Gamma}(V')+\int_{max(0,\Gamma-V)}^{min(\Gamma,V)} dV' 
 \psi_{\Gamma}(V') \right)
\end{eqnarray}

We now differentiate this equation with respect to $V$
\begin{eqnarray} 
&& \left[ \partial_{\Gamma}+(V-\Gamma) n_{\Gamma} \right] \partial_V \psi_{\Gamma}(V) =0
\ \ \ \ \   \hbox{for}  \ \  V \geq \Gamma  \\
&&   \partial_{\Gamma}  \partial_V \psi_{\Gamma}(V) =
2 n_{\Gamma} \left(\psi_{\Gamma}(V) + \psi_{\Gamma}(\Gamma-V) \right)
\ \ \ \ \   \hbox{for}  \ \  0 \leq V \leq \Gamma 
\label{eqpsi}
\end{eqnarray}
and keep the following condition at $V=\Gamma$
\begin{eqnarray} 
  \partial_{\Gamma} \psi_{\Gamma}(V) \vert_{V=\Gamma}
= 2  n_{\Gamma}  \int_{0}^{\Gamma} dV' 
 \psi_{\Gamma}(V')
\label{conditionveqgamma}
\end{eqnarray}

The normalisation condition (\ref{normadrift}) now reads
\begin{eqnarray} 
&& \int_0^F dW D^{\pm}_{\Gamma}(F,W)=
n_{\Gamma}
\int_0^{\infty} dl l P^{\pm}_{\Gamma}(F,l) 
 =- n_{\Gamma}
\partial_p \left( U^{\pm}_{\Gamma}(p) e^{- (F-\Gamma) u^{\pm}_{\Gamma}(p) } \right) \vert_{p=0} \\
&& = n_{\Gamma} 
u^{\pm}_{\Gamma}(0) e^{- (F-\Gamma) u^{\pm}_{\Gamma}(0) }
\left( - \partial_p \ln U^{\pm}_{\Gamma}(p) \vert_{p=0}
+(F-\Gamma) \partial_p  u^{\pm}_{\Gamma}(p) \vert_{p=0}   \right)
\end{eqnarray}
Taking into account the changes of variables
\begin{eqnarray} 
&& \int_0^F dW D^{\pm}_{\Gamma}(F,W)
= n_{\Gamma} u^{\pm}_{\Gamma}(0) e^{- (F-\Gamma) u^{\pm}_{\Gamma}(0) }
\int_0^F dW \Psi_{\Gamma}^{\pm}(F-W,W) \\
&& = 2 n_{\Gamma} u^{\pm}_{\Gamma}(0) e^{- (F-\Gamma) u^{\pm}_{\Gamma}(0) }
\int_0^F dW \psi_{\Gamma}(W)
\end{eqnarray}
we get for $F \geq \Gamma$
\begin{eqnarray} 
 2 \int_0^F dW \psi_{\Gamma}(W)
=  - \partial_p \ln U^{\pm}_{\Gamma}(p) \vert_{p=0}
+(F-\Gamma) \partial_p \ln u^{\pm}_{\Gamma}(p) \vert_{p=0}  
\end{eqnarray}

Deriving with respect to $F$, we get for any $F \geq \Gamma$
\begin{eqnarray} 
    \psi_{\Gamma}(F \geq \Gamma)=\psi_{\Gamma}( \Gamma)
=\frac{1}{2} \partial_p  u^{\pm}_{\Gamma}(p) \vert_{p=0}
=\frac{ (\sinh (2 \delta \Gamma) -2 \delta \Gamma)}
{ 8 \delta \sinh^2( \delta \Gamma)}
\label{psigamma}
\end{eqnarray}

For the value $F=\Gamma$ we get

\begin{eqnarray} 
\int_0^{\Gamma} dW \psi_{\Gamma}(W)
= - \frac{1}{2} \partial_p \ln U^{\pm}_{\Gamma}(p) \vert_{p=0}
= \frac{ \delta \Gamma \coth ( \delta \Gamma) - 1}
{ 4  \delta^2 }
\end{eqnarray}

The equation (\ref{eqpsi}) for $V \geq \Gamma$ is thus satisfied, as 
well as the condition (\ref{conditionveqgamma}) at $V=\Gamma$. 
We now have to find the solution $\psi_{\Gamma}(V)$ 
to the equation (\ref{eqpsi}) on the domain $0 \leq V \leq \Gamma$
that satisfies the continuity boundary conditions
$\psi_{\Gamma}(V=\Gamma)=\psi_{\Gamma}(\Gamma)$
given in (\ref{psigamma})
 and $\partial_V \psi_{\Gamma}(V) \vert_{V=\Gamma} =0$.
It reads
\begin{eqnarray} 
\psi_{\Gamma}(V)
=\psi_{\Gamma}(\Gamma)-
\frac{ (\sinh (2 \delta (\Gamma-V)) -2 \delta (\Gamma-V))}
{ 4 \delta \sinh^2( \delta \Gamma)}
\end{eqnarray}

The final result is thus that the joint distributions $D_{\Gamma}^{\pm}(F,W)$
are given by
\begin{eqnarray} 
 D_{\Gamma}^{\pm}(F,W) 
=n_{\Gamma} u^{\pm}_{\Gamma} e^{- (F-\Gamma) u^{\pm}_{\Gamma}}
\left( \psi_{\Gamma}(W) + \psi_{\Gamma}(F-W)\right) 
\end{eqnarray}
where 
\begin{eqnarray} 
 \psi_{\Gamma}(W)= \frac{1}{2} h_{\Gamma}(\Gamma) - \theta(\Gamma-W)  h_{\Gamma}(\Gamma-W)
\end{eqnarray}
in terms of the auxiliary function
\begin{eqnarray} 
 h_{\Gamma}(V)=\frac{ (\sinh (2 \delta V) -2 \delta V)}
{ 4 \delta \sinh^2( \delta \Gamma)}
\end{eqnarray}

We now compute the distributions of $W$ alone
\begin{eqnarray} 
&& D_{\Gamma}^{\pm}(W) 
 = \int_{\hbox{max}(\Gamma,W)}^{+\infty} dF D_{\Gamma}^{\pm}(F,W) \\
&& = 
 n_{\Gamma} 
\left( h_{\Gamma}(\Gamma)
- \theta(\Gamma-W)  h_{\Gamma}(\Gamma-W) \right)  
 e^{- \hbox{ max}(0,W-\Gamma) u^{\pm}_{\Gamma}}  - n_{\Gamma} \int^{\hbox{min}(\Gamma,W)}_{0} dV
 u^{\pm}_{\Gamma} e^{(V-W) u^{\pm}_{\Gamma}} h_{\Gamma}(V) 
\end{eqnarray}

More explicitly, for $W \geq \Gamma$, we have a simple exponential form in $W$
\begin{eqnarray} 
 D_{\Gamma}^{\pm}(W>\Gamma) 
 && = e^{- (W-\Gamma) u^{\pm}_{\Gamma}}
 n_{\Gamma} \left[
 h_{\Gamma}(\Gamma) 
 -  u^{\pm}_{\Gamma} \int^{\Gamma}_{0} dV
  e^{-(\Gamma-V) u^{\pm}_{\Gamma}} h_{\Gamma}(V) \right)
\end{eqnarray}
whereas for $0 \leq W \leq \Gamma$, we have
\begin{eqnarray} 
&& D_{\Gamma}^{\pm}(W<\Gamma) 
 =  n_{\Gamma} 
\left( h_{\Gamma}(\Gamma)
-   h_{\Gamma}(\Gamma-W)   
   -   u^{\pm}_{\Gamma} \int^{W}_{0} dV
 e^{(V-W) u^{\pm}_{\Gamma}} h_{\Gamma}(V) \right)
\end{eqnarray}

These expressions lead to the final results (\ref{rgbiaswg1})
and (\ref{rgbiasws1}) given in the text.

\subsection{Joint distribution of position and energy }

\label{jointbias}

The path-integral approach to compute the
joint distribution of position and energy
can be generalized to the biased case as follows.

The basic path-integral (\ref{fpath}) becomes
\begin{eqnarray}
F_{[V_a,V_b]}^{\pm}(V_0, x_0, V , x ) && = 
\int_{V_{x_0}=V_0}^{V_{l}=V} DV(y) 
exp( - \int_0^x dy ( \frac{1}{4} (\frac{dV}{dy} \mp  2 \delta)^2  )
\theta_{[V_a,V_b]}^-(\{V\}) \\
&& = e^{- \delta^2 (x-x_0) \pm \delta (V-V_0) } F_{[V_a,V_b]}(V_0, x_0, V , x )
\end{eqnarray}
and thus in Laplace transform with respect to $(x-x_0)$,
we have in terms of (\ref{res-system})
\begin{eqnarray}
{\hat F}_{[V_a,V_b]}^{\pm}(V,p \vert V_0)  = 
  e^{ \pm \delta (V-V_0) } 
{\hat F}_{[V_a,V_b]}(V,p+\delta^2 \vert V_0)
\end{eqnarray}

Similarly, (\ref{resB-sym}) and (\ref{resE}) become
\begin{eqnarray} 
{\hat B}_{\Gamma}^{\pm}(F,p )
&&  = e^{\pm \delta \Gamma} \frac{ {\sqrt {p+\delta^2} } }{  \sinh {\sqrt {p+\delta^2}} \Gamma }
 e^{- (F-\Gamma) \left(\mp \delta+ {\sqrt {p+\delta^2}}  \coth {\sqrt {p+\delta^2}} \Gamma \right)  }
 \\
{\hat E}_{\Gamma}(V,p \vert V_0 )
&&  =  e^{- (V-V_0) \left(\mp \delta+ {\sqrt {p+\delta^2}}  \coth {\sqrt {p+\delta^2}} \Gamma \right) }
\end{eqnarray}

\subsubsection{ Case $W>\Gamma$  }

Using the same decomposition as in the symmetric case (see Figure \ref{fig1}
and (\ref{decomposition})),
the path-integral (\ref{jointpathbiased}) becomes

\begin{eqnarray} 
{\hat P}_t^{\pm} (x, W)&& =  \frac{\delta^2}{\sinh^2 \delta \Gamma}  
 \int_W^{W+\Gamma} dV_m \int_0^x dx_m B_{\Gamma}^{\pm} (V_m,x_m) 
\left (\partial_{V_0} F_{[V_m-\Gamma,V_m]}^{\pm}(W,x\vert V_0,x_m ) \right) \vert_{V_0=V_m} \\
&& \int_x^{+\infty} dx_m' 
\left(\partial_{V} F_{[V_m-\Gamma,V_m]}^{\pm}(W,x_m'\vert V,x )
\right)\vert_{V=V_m}
\int_{V_m}^{+\infty} dF \int_{x_m'}^{+\infty} dl E_{\Gamma}^{\pm} (F,l ; V_m,x_m')
\label{decompobias}
\end{eqnarray}
and thus we obtain in Laplace transform 
\begin{eqnarray} 
{\hat P}_t^{\pm} (p, W) && =  \frac{\delta^2}{\sinh^2 \delta \Gamma}
\int_W^{W+\Gamma} dV_m e^{  \pm \delta \Gamma }\frac{ {\sqrt {p +\delta^2}} }{  \sinh {\sqrt  {p +\delta^2}} \Gamma }
 e^{- (V_m-\Gamma) ( \mp \delta +{\sqrt  {p +\delta^2}}  \coth {\sqrt  {p +\delta^2}} \Gamma)}
\\
&& \frac{\sinh {\sqrt {p+\delta^2}} (\Gamma-(V_m- W))}{\sinh {\sqrt {p+\delta^2}} \Gamma}  \frac{\sinh \delta (\Gamma-(V_m- W))}{\sinh \delta \Gamma}
\int_{V_m}^{+\infty} dF 
 e^{- (F-V_m) (\mp \delta+\delta \coth \delta \Gamma ) } 
\end{eqnarray}
which leads to the final result (\ref{biasjointplus}) given in the text.

\subsubsection{ Case $W<\Gamma$  }

As in the symmetric case, there are two contributions.
The first one corresponding to the case where $V_m>\Gamma$
is the same as (\ref{decompobias}) except that now we have to integrate
over  $V_m \in (\Gamma,W+\Gamma)$ (instead of $V_m \in (W,W+\Gamma)$)
and we get

\begin{eqnarray} 
&& {\hat P}_t^{(1)\pm} (p, W<\Gamma) 
 =  \frac{\delta^2}{\sinh^2 \delta \Gamma}
\int_{\Gamma}^{W+\Gamma} dV_m e^{  \mp \delta \Gamma }\frac{ {\sqrt {p +\delta^2}} }{  \sinh {\sqrt  {p +\delta^2}} \Gamma }
 e^{- (V_m-\Gamma) ( \pm \delta +{\sqrt  {p +\delta^2}}  \coth {\sqrt  {p +\delta^2}} \Gamma)}  \\
&& \frac{\sinh {\sqrt {p+\delta^2}} (\Gamma-(V_m- W))}{\sinh {\sqrt {p+\delta^2}} \Gamma}  \frac{\sinh \delta (\Gamma-(V_m- W))}{\sinh \delta \Gamma}
\int_{V_m}^{+\infty} dF 
 e^{- (F-V_m) (\pm \delta+\delta \coth \delta \Gamma ) }  \\
&& = \frac{\delta^2}{\sinh^2 \delta \Gamma \left(
\frac{p+\delta^2}{\sinh^2 {\sqrt  {p +\delta^2}} \Gamma} \mp 4 \delta ( \mp \delta + {\sqrt  {p +\delta^2}}  \coth {\sqrt  {p +\delta^2}} \Gamma)  \right) } \\
&& [-2   ( \mp \delta +{\sqrt  {p +\delta^2}}  \coth {\sqrt  {p +\delta^2}} \Gamma)
 e^{-W ( \mp \delta +{\sqrt  {p +\delta^2}}  \coth {\sqrt  {p +\delta^2}} \Gamma)} \\
&& +2 ( \mp \delta +{\sqrt  {p +\delta^2}}  \coth {\sqrt  {p +\delta^2}} \Gamma) \cosh {\sqrt  {p +\delta^2}} W 
\cosh \delta W \\
&& -  ({\sqrt  {p +\delta^2}}(1+\coth^2 {\sqrt {p+\delta^2}} \Gamma)
\mp 2 \delta \coth {\sqrt  {p +\delta^2}} \Gamma) ) 
\sinh {\sqrt  {p +\delta^2}}  W \cosh \delta W \\
&& + \frac{1}{\delta} (p-( \mp \delta +{\sqrt  {p +\delta^2}}  \coth {\sqrt  {p +\delta^2}} \Gamma)^2)\cosh {\sqrt  {p +\delta^2}}  W \sinh \delta W  \\
&& +  ( \mp \delta +{\sqrt  {p +\delta^2}}  \coth {\sqrt  {p +\delta^2}} \Gamma )(-2  \coth {\sqrt  {p +\delta^2}} \Gamma +
\frac{{\sqrt {p+\delta^2}}}{ \delta \sinh^2 {\sqrt {p+\delta^2}} \Gamma})\sinh {\sqrt  {p +\delta^2}}  W \sinh \delta W]
\label{biasmoinscontri1}
 \end{eqnarray}

The second contribution corresponding to the case where $V_m<\Gamma$
(see Figure (\ref{fig2}) can be obtained as in (\ref{decompomoinssym}) 
\begin{eqnarray} 
{\hat P}^{(2)\pm}_t (x, W<\Gamma)&& = 
 \frac{\delta^2}{\sinh^2 \delta \Gamma} 
\left(\partial_{V_0} F_{[0,\Gamma]}^{\pm}(V_0,0;W,x) \right)_{V_0=0} 
\int_x^{\infty} dz \left(\partial_{V'} F_{[0,\Gamma]}^{\pm}(W,x;,V',z) \right)_{V'=\Gamma} \\
&& \int_{\Gamma}^{+\infty} dF \int_z^{+\infty} dl E_{\Gamma}^{\pm}(\Gamma,z;F,l) \end{eqnarray}
and thus in Laplace transform

\begin{eqnarray} 
{\hat P}_t^{(2)\pm} (p, W) 
&& =  \frac{\delta^2}{\sinh^2 \delta \Gamma}
\left(\partial_{V_0} F_{[0,\Gamma]}^{\pm}(w,p \vert V_0) \right)_{V_0=0} 
\int_x^{\infty} dz \left(\partial_{V'} F_{[0,\Gamma]}^{\pm}(V',0 \vert W) \right)_{V'=\Gamma} \\
&& \int_{\Gamma}^{+\infty} dF  E_{\Gamma}^{\pm}(F,0 \vert \Gamma) \\
&& = \frac{\delta \sinh \delta W \sinh (\Gamma-W) {\sqrt {p+\delta^2}} }
{\sinh^2 \delta \Gamma \sinh \Gamma {\sqrt {p+\delta^2}}}
\label{biasmoinscontri2}
\end{eqnarray}

The final result for $W<\Gamma$ is obtained by summing the two contributions
(\ref{biasmoinscontri1}) and (\ref{biasmoinscontri2})
and is given in (\ref{biasjointmoins}) in the text.


\begin{thebibliography}{10}


\bibitem{review}
see for reviews, e.g. J.P. Bouchaud et al. in 
''Spin glasses and random fields'', A.P. Young ed
World Scientific Singapore 1998, 
G. Blatter et al. Rev. Mod. Phys. {\bf 66} 1125 (1994).

\bibitem{frg}
L. Balents, P. Le Doussal, cond-mat/0205358
Chauve et al. cond-mat/0002299 

\bibitem{dilute}
G. Semerjian, L. F. Cugliandolo, cond-mat/0204613

\bibitem{sinai}
Y. G. Sinai Theor. Probab. Its Appl. {\bf 27} 247 (1982).

\bibitem{kesten}
H. Kesten, Physica {\bf 138} A 299 (1986).

\bibitem{derrida_pomeau}
B. Derrida, J. Stat. Phys. {\bf 31} 433 (1983).

\bibitem{ledou_1d}
J. P. Bouchaud, A. Comtet, A. Georges, P. Le Doussal Europhys. Lett. {\bf 3}
  653 (1987), Ann Phys. {\bf 201} 285 (1990).

\bibitem{monthus_diffusion}
A. Comtet, J. Desbois and C. Monthus Ann. Phys. {\bf 239} 312 (1995). C.
  Monthus et al. Phys. Rev. E {\bf 54} 231 (1996).

\bibitem{laloux_pld_sinai}
L. Laloux and P. Le Doussal cond-mat/9705249, Phys. Rev. E {\bf 57} 6296 (1998)
  and references therein.

\bibitem{us_prl}
D.S. Fisher, P. Le Doussal and C. Monthus, Phys. Rev. Lett.
  {\bf 80} 3539 (1998).

\bibitem{us_sinai}
 P. Le Doussal, C. Monthus, D.S. Fisher,  Phys. Rev. E {\bf 59} 4795
(1999)


\bibitem{ma_dasgupta}
C. Dasgupta and S.K. Ma Phys. Rev. B {\bf 22} 1305 (1980).

\bibitem{daniel}
D.S. Fisher, Phys. Rev. B {\bf 50} 3799 (1994);
D.S. Fisher, Phys. Rev. B {\bf 51} 6411 (1995).

\bibitem{us_golosov}
C. Monthus and P. Le Doussal, cond-mat/0202295.

\bibitem{us_toy}
 P. Le Doussal and C. Monthus, cond-mat/0204168.

\bibitem{us_rfim}
Daniel S. Fisher, Pierre Le Doussal, Cecile Monthus,
cond-mat/0012290

\bibitem{us_rd}
Pierre Le Doussal, Cecile Monthus, cond-mat/9901306

\bibitem{bounds}
L. Cugliandolo et al. cond-mat/9705002 

\end{thebibliography}

\end{document}